\documentclass[pre,twocolumn,showpacs,superscriptaddress]{revtex4}

\usepackage{graphicx}
\usepackage{amsmath}
\usepackage{color}
	\usepackage[colorlinks,hyperindex]{hyperref}
 \usepackage{hyperref}
	\hypersetup
	{
		colorlinks,%
		citecolor=blue,%
		linkcolor=blue,%
		urlcolor=black,%
	}	

\begin{document}

%===================================================================
\title{Wrinkles and folds in a fluid-supported sheet of finite size}
%===================================================================

\author{Oz Oshri} 
\email{ozzoshri@tau.ac.il} 
\affiliation{Raymond \& Beverly Sackler School of Physics \& Astronomy, Tel Aviv University, Tel Aviv 6997801, Israel}

\author{Fabian Brau}
\email{fabian.brau@ulb.ac.be}
\affiliation{Nonlinear Physical Chemistry Unit, Universit\'e libre de Bruxelles (ULB), CP231, 1050 Brussels, Belgium}

\author{Haim Diamant} 
\email{hdiamant@tau.ac.il} 
\affiliation{Raymond \& Beverly Sackler School of Chemistry, Tel Aviv University, Tel Aviv 6997801, Israel}

\date{\today}

\begin{abstract} 
A laterally confined thin elastic sheet lying on a liquid substrate displays regular undulations, called wrinkles, characterized by a spatially extended energy distribution and a well-defined wavelength $\lambda$. As the confinement increases, the deformation energy is progressively localized into a single narrow fold. An exact solution for the deformation of an infinite sheet was previously found, indicating that wrinkles in an infinite sheet are unstable against localization for arbitrarily small confinement. We present an extension of the theory to sheets of finite length $L$, accounting for the experimentally observed wrinkle-to-fold transition. We derive an exact solution for the periodic deformation in the wrinkled state, and an approximate solution for the localized, folded state. We find that a second-order transition between these two states occurs at a critical confinement $\Delta_{\text{F}} = \lambda^2/L$.
\end{abstract}


\maketitle
%------------------------------------------------

\section{Introduction}
%---------------------
\label{sec_intro}

Morphological transitions are often induced by confinement or by spatially constrained growth. Structures emerge spontaneously when the energy injected into a system through the confinement process ceases to distribute uniformly. These phenomena are observed in various contexts ranging from the folding of geological layers~\cite{hudl10} to patterns in biological membranes and
monolayers~\cite{miln89,kaga99,lu02,jpc06,zhang07,lee08} and the formation of fingerprints~\cite{kuck04,kuck05}. Besides the initial morphology, occurring for small confinement, various subsequent transitions may be observed as the confinement increases.

\begin{figure}
\includegraphics[width=\columnwidth]{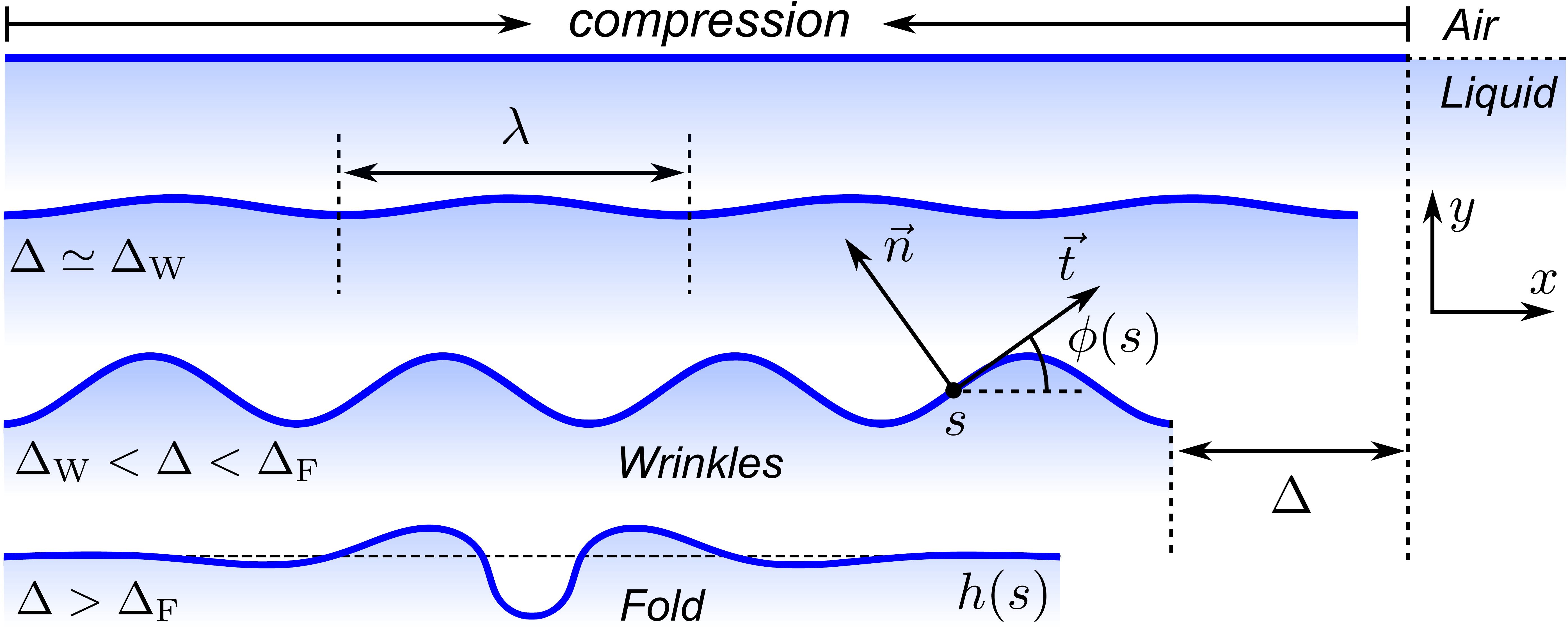}
\caption{(Color online) Schematic morphological evolution of a compressed finite sheet floating on a liquid. The scheme shows the emergence of a wavelength $\lambda$ upon small confinement, $\Delta \gtrsim \Delta_{\text{W}}$, the growth in amplitude of the wrinkle state for intermediate compression, and the transition to a fold state beyond some critical confinement, $\Delta_{\text{F}}$. Here $\vec{t}$ and $\vec{n}$ are the tangent and normal to the sheet surface, respectively. $\phi(s)$ is the angle between the local tangent and the horizontal direction $x$, and the arclength along the sheet is parametrized by $s$. $h(s)$ is the vertical elevation of the sheet.}
\label{fig01}
\end{figure}

Many model systems have been developed to study the influence of confinement on morphological transitions~\cite{cerda03,cerda04,benny09,reis09,holmes10,ebata12,li12} using rods~\cite{dona02,stoo08,baya11} or sheets resting on some substrate~\cite{hunt93,lee96,boue06,debo09,ahar10,huan10,brau11,camb11,schr11,vand11,seml13,lijun15}. Among them, the experimental model system of a thin elastic sheet, lying on a fluid substrate~\cite{vella04,huang07,leahy10,vella10,wagner11,king12,rivetti13,schroll13,pine13,deme14} and subjected to in-plane uniaxial compression, deserves special interest~\cite{poci08,arxiv10,audo11,diam11,brau13,diam13,rive13,rive14}. Beyond a certain critical confinement, $\Delta_{\text{W}}$, which vanishes for an incompressible sheet, the sheet buckles, displaying regular undulations with a characteristic wavelength $\lambda$ over its entire length~\cite{miln89,poci08} (Fig.~\ref{fig01}). Beyond a certain confinement, $\Delta_{\text{F}}$, another transition occurs, where the wrinkles start attenuating except near the center of the sheet, and the deformation energy gets localized into a single fold~\cite{poci08}.

In the case of an infinitely long sheet, an exact solution describing the complete evolution of the morphology with increasing confinement proves that such a wrinkle-to-fold transition, strictly, should never occur~\cite{diam11}, \textit{i.e.} $\Delta_{\text{F}}\rightarrow\Delta_{\text{W}}$. The infinite-sheet morphology is always localized, but the localization decay length
diverges as the buckling threshold $\Delta_{\text{W}}$ is approached.

Even if an infinitely long sheet is a useful idealization of real
systems allowing a good agreement with experiment~\cite{brau13}, the
apparent discrepancy between theory and experiment concerning the
existence of a wrinkle-to-fold transition should be
resolved. Experiments are obviously performed with finite sheets. We
demonstrate that a second-order wrinkle-to-fold transition does occur
for finite sheets. In Sec.~\ref{sec_wrinkle} we construct an exact
periodic solution for wrinkles of arbitrary amplitude in a finite
sheet. An exact solution for a localized deformation in a finite sheet
has evaded us. Hence, in Sec.~\ref{sec_fold}, we derive an approximate
localized solution using a multiple-scale analysis. In
Sec.~\ref{sec_transition}, we identify the order parameter of the
transition in the context of the Landau theory of second-order
transitions. We show how both periodic and localized solutions match
at the critical confinement, $\Delta_{\text{F}} = \lambda^2/L$. In
Sec.~\ref{sec_discuss}, we discuss the experimental implications of
our theory, its limitations, and future extensions.

\section{System and governing equation}
%-------------------
\label{sec_system}

The system studied here is composed of a thin incompressible elastic sheet ($\Delta_{\text{W}}=0$) of length $L$, width $W$, and bending modulus $B$, lying on a fluid of mass density $\rho$. The sheet is uniaxially confined by a distance $\Delta$ along the $x$-axis and deforms in the $xy$ plane. The shape of the sheet is described by the parametric equation 
\begin{subequations}
\label{param-eq-xh}
\begin{align}
\label{param-eq-x}
x(s) &= \int_{-L/2}^s \cos\phi(s') ds', \\
\label{param-eq-h}
h(s) &= \int_{-L/2}^s \sin\phi(s') ds',
\end{align}
\end{subequations}
where $\phi(s)$ is the angle between the local tangent to the sheet and the $x$-axis at a given arclength $s$, see Fig.~\ref{fig01}. The total energy of the system, $E$, is composed of the bending energy of the sheet, $E_{\text{b}}=(WB/2)\int_{-L/2}^{L/2} \dot{\phi}^2 ds$, and the deformation energy of the substrate, $E_{\text{s}}=(W\rho g/2) \int_{-L/2}^{L/2} h^2 \cos \phi\, ds$, where the dot denotes an $s$ derivative~\cite{poci08,diam11}. The displacement along the direction of confinement is given by
\begin{equation}
\label{delta-phi}
\Delta = \int_{-L/2}^{L/2} (1-\cos \phi)\, ds,
\end{equation}
and is related to the applied load necessary to confine the sheet by $P = dE/d\Delta$. In the following, except where it is explicitly mentioned, we use units such that the energy is rescaled by $B$, and lengths are rescaled by $(B/\rho g)^{1/4}\equiv \lambda/2\pi$. As a result, the applied load $P$ is rescaled by $(B \rho g)^{1/2}$.

To find the equilibrium shape, one should minimize the total energy under appropriate constraints. This can be reformulated as a dynamical problem \cite{diam11} with an action ${\cal S}=
\int_{-L/2}^{L/2} {\cal L} (\phi,h,\dot{\phi},\dot{h})$, where
\begin{equation}
\label{lagrangian}
{\cal L} = \frac{1}{2} \dot{\phi}^2 + \frac{1}{2} h^2 \cos\phi-P(1-\cos\phi -\Delta/L)-Q(s)(\sin\phi -\dot{h}).
\end{equation}
In Eq.~(\ref{lagrangian}), $P$ and $Q(s)$ are Lagrange multipliers introduced to take into account, respectively, the global constraint (\ref{delta-phi}) and the local geometrical constraint between $h$ and $\phi$. The conjugate momenta are defined as $p_{\phi} = \partial {\cal L}/\partial \dot{\phi} = \dot{\phi}$ and $p_{h} = \partial {\cal L}/\partial \dot{h} = Q$ and are used to construct the Hamiltonian ${\cal H} = p_{\phi}\dot{\phi} + p_h \dot{h} - {\cal L}$. Since ${\cal L}$ has no explicit dependence on $s$, the Hamiltonian is a constant for a given displacement $\Delta$,
\begin{equation}
\label{hamiltonian}
{\cal H}=\frac{1}{2}p_{\phi}^2 + p_h \sin \phi - \frac{1}{2} h^2 \cos\phi + P (1-\cos\phi -\Delta/L) = C.
\end{equation}
Hamilton's equation, $\dot{p}_{\phi}=-\partial {\cal H}/\partial \phi$, yields the following equation:
\begin{equation}
\label{eq-motion}
\ddot{\phi} + (h^2/2 + P)\sin\phi +p_h \cos\phi = 0.
\end{equation}
Eliminating $p_h$ between Eqs.~(\ref{hamiltonian}) and (\ref{eq-motion}), we obtain
\begin{equation}
\label{eq-motion2}
\ddot{\phi} \sin\phi - \frac{1}{2} \dot{\phi}^2 \cos\phi + \frac{1}{2} h^2 + P -\tilde{P} \cos \phi=0,
\end{equation}
where $\tilde{P} \equiv P(1-\Delta/L)-C$ is a shifted value of the load, dependent on boundary conditions. Differentiating Eq.~(\ref{eq-motion2}) once, we obtain the well known equation for Euler's elastica,
\begin{equation}
\label{elastica-hydro}
\dddot{\phi}+\left(\frac{1}{2}\dot{\phi}^2+\tilde{P}\right)\dot{\phi}+h=0,
\end{equation} 
where the local normal force exerted on the elastica is given by the hydrostatic term $h$. A second differentiation gives the equation governing the system evolution,
\begin{equation}
\label{main-equation}
\ddddot{\phi} + \frac{3}{2} \dot{\phi}^2 \ddot{\phi} + \tilde{P} \ddot{\phi} + \sin \phi =0.
\end{equation}
Finally, Eq.~(\ref{eq-motion2}) is used to relate the physical load $P$ to the shifted value $\tilde{P}$. For hinged boundary conditions, where $h$ and $\ddot{h}$ (and also $\dot{\phi}$) vanish at the boundaries, we have
\begin{equation}
\label{P-hinged}
P = \tilde{P}\cos \phi(\pm L/2) - \ddot{\phi}(\pm L/2)\sin\phi(\pm L/2).
\end{equation}
As a result, once the solution $\phi(s)$ is obtained, the physical load $P$ can be computed from $\tilde{P}$. Note that in the limit of an infinite sheet, where $\phi(\pm L/2)\rightarrow 0$, $P$ and $\tilde{P}$ coincide. 

The total rescaled energy per unit length of the system reads
\begin{equation}
\label{ener-phi}
E[\phi(s)] = \frac{1}{2} \int_{-L/2}^{L/2} ds \left(\dot{\phi}^2 + h^2 \cos\phi \right).
\end{equation}
The energy, the displacement $\Delta$ and the equation giving the equilibrium shape of the sheet can also be written in terms of $h$ instead of $\phi$ using $\dot{h}=\sin\phi$:
\begin{eqnarray}
\label{ener-h}
E[h(s)] &=& \frac{1}{2} \int_{-L/2}^{L/2} ds \left(\frac{\ddot{h}^2}{1-\dot{h}^2} + h^2\sqrt{1-\dot{h}^2} \right), \\
\label{delta-h}
\Delta[h(s)] &=& \int_{-L/2}^{L/2} ds \left( 1-\sqrt{1-\dot{h}^2} \right),
\end{eqnarray}
The shape of the sheet is determined from minimization of either $E$, given the displacement $\Delta$, or the function,
\begin{equation}
\label{def-g}
G = E - P\Delta, 
\end{equation}
given the load $P$.

\section{Exact periodic solutions}

\label{sec_wrinkle}

\subsection{General results}

We first study general solutions of Eq.~(\ref{main-equation}) without specifying the boundary conditions. In order to construct a periodic solution of Eq.~(\ref{main-equation}), we recall the connection existing between this equation and the dynamics of a physical pendulum~\cite{gesz03,diam11}. For this purpose, we consider the total energy $U_{\text{p}}$ of a pendulum,
\begin{equation}
\label{energy}
U_{\text{p}}\equiv 2k^2 q^2 = \frac{\dot{\phi}^2}{2}+q^2(1-\cos \phi),
\end{equation}
where $q$ is the natural frequency of the pendulum, and $k$ is a constant determined by the pendulum's total energy or, equivalently, by boundary conditions. (In the analogous elastic sheet these two parameters are related to the natural wrinkling wavenumber, $2\pi/\lambda$, and the total displacement $\Delta$.) The equation of motion is obtained by taking the variation of $U_{\rm p}$ in the above equation,
\begin{equation}
\label{pendule}
\ddot{\phi} + q^2 \sin \phi =0.
\end{equation}
Differentiating twice this last equation and eliminating $q^2 \cos \phi$ and $q^2 \sin \phi$ using respectively Eqs.~(\ref{energy}) and (\ref{pendule}), we obtain
\begin{equation}
\ddddot{\phi} + \frac{3}{2} \dot{\phi}^2 \ddot{\phi} + \left(1-2k^2\right)q^2 \ddot{\phi}=0.
\end{equation}
Adding Eq.~(\ref{pendule}) multiplied by $q^{-2}$, we finally obtain the equation describing the morphology of a confined floating sheet,  see Eq.~(\ref{main-equation}):
\begin{subequations}
\begin{align}
\label{membrane-eq}
\ddddot{\phi} &+ \frac{3}{2} \dot{\phi}^2 \ddot{\phi} + \tilde{P} \ddot{\phi} + \sin \phi =0, \\
\label{eq-p0}
\tilde{P} &= q^2 \left(1-2k^2\right)+q^{-2}.
\end{align}
\end{subequations}
Therefore, any solution of Eq.~(\ref{pendule}) is also a solution of Eq.~(\ref{main-equation}) with $\tilde{P}$ given by Eq.~(\ref{eq-p0}).

It is well known that Eq.~(\ref{pendule}) admits a periodic solution for $0\leq k<1$ in terms of Jacobian elliptic functions~\cite[p.549]{olve10},\footnote{Note that there are two definitions for the modulus of the elliptic functions. In this section, we use the one given in Ref.~\cite{olve10}. The other one, $m=k^2$, is used in Ref.~\cite{Abramowitz} and by {\it Mathematica}\copyright.}:
\begin{equation}
\label{sol-pendule}
\phi(s) = 2 \arcsin\left[k\, \text{sn}(q (s+s_0), k) \right].
\end{equation}
The profile $h(s)=\int_{-L/2}^{s} \sin \phi(s') ds'$ is thus given by
\begin{equation}
\label{profile-period}
h(s)=\frac{2k}{q} \left[\text{cn}(q(s+s_0),k)-\text{cn}(q(-L/2+s_0),k)\right].
\end{equation}
This solution has a periodicity of $4K(k)/q$, where $K(x)$ is the complete elliptic integral of the first kind~\cite[p.487]{olve10}. Finally, the horizontal displacement is given by
\begin{align}
\label{delta-period}
\Delta &= 2k^2\int_{-L/2}^{L/2} \text{sn}^2(q(s+s_0),k)\, ds \nonumber \\
&= 2L - \frac{2}{q}\left\{{\cal E}(q(s+s_0),k)\right\}_{-L/2}^{L/2},
\end{align}
where ${\cal E}(x,k)$ is the Jacobi epsilon function~\cite[p.562]{olve10}. The quantities $s_0$ and $q$ are fixed to satisfy the boundary conditions whereas the parameter $k$ is related to the confinement ratio, $\Delta/L$.

\subsection{Hinged sheets}

For hinged boundary conditions both $h$ and $\ddot{h}$ vanish at $s=\pm L/2$. The second derivative of the sheet profile, Eq.~(\ref{profile-period}), is given by
\begin{equation}
\label{ypp-period}
\ddot{h}(s) = 2kq \, \text{cn}(q(s+s_0),k) \left(2\,\text{dn}^2(q(s+s_0),k)-1 \right).
\end{equation}
The $\text{dn}$ function varies between $(1-k^2)^{1/2}$ and $1$; consequently, so long as $k<1/\sqrt{2}$, the last factor in Eq.~(\ref{ypp-period}) does not vanish. As shown below, the relevant values of $k$ are smaller than $1/3$ provided $L \ge 3\lambda$. Larger values of $k$ lead to periodic solutions unstable against localization.

Due to the periodicity of the solution, there exists an infinite number of possible solutions depending on the number of nodes. From Eq.~(\ref{profile-period}), $h(L/2)=0$ provided that $qL/2+qs_0 = -qL/2+qs_0 + 4n_1 K(k)$. From Eq.~(\ref{ypp-period}), $\ddot{h}(L/2)=0$ provided that $qL/2 +qs_0 = (2n_2+1) K(k)$, with $n_1$ and $n_2$ positive integers. Due to the definite parity of the solution (either symmetric or antisymmetric), the condition $\ddot{h}(-L/2)=0$ is then automatically satisfied. We have the following two possibilities. (a) If $s_0=0$, then $h(L/2)=0$ automatically, for any $q$, due to the even parity of the $\text{cn}$ function in Eq.~(\ref{profile-period}). We are left with the condition for $\ddot{h}(L/2)=0$ which gives $qL=2(2n_2+1)K(k)$. (b) If $s_0\neq 0$, then the two conditions above must be satisfied simultaneously, giving $qL=4n_1 K(k)$ and $q s_0=[2(n_2-n_1)+1]K(k)$. Thus, combining these two results, we have
\begin{align}
\label{q-s0}
q &= \frac{2(2p+1) K(k)}{L} \quad \text{with} \quad s_0=0 \\
q &= \frac{2(2p) K(k)}{L} \quad \text{with} \quad qs_0= K(k),
\end{align}
where $p$ is a strictly positive integer. Finally, the solution reads
\begin{subequations}
\begin{align}
\label{phi-profile-period-hinged}
\phi(s) &= 2 \arcsin\left[k\, \text{sn}(qs + \varepsilon K(k),k)\right], \\
\label{profile-period-hinged}
h(s)&= \frac{2k}{q} \text{cn}(qs + \varepsilon K(k) ,k),
\end{align}
\end{subequations}
where $\varepsilon = (1+(-1)^n)/2$ and
\begin{equation}
\label{def-q}
q = \frac{2n K(k)}{L}, \quad n=1,2,3 \ldots 
\end{equation}
Symmetric solutions correspond to $n$ of odd parity and antisymmetric solutions are obtained with an even parity of $n$. The number of nodes of the solution is equal to $n+1$ (counting the two nodes at the boundaries). 

\subsection{Pressure, displacement, and amplitude}
\label{sec_pressure_period}

From Eq.~(\ref{phi-profile-period-hinged}), we have 
\begin{align}
\cos\phi(L/2) &= 1-2k^2, \quad \ddot{\phi}(L/2) = \mp 2k q^2 \sqrt{1-k^2} \nonumber \\
\sin\phi(L/2) &= \pm 2 k \sqrt{1-k^2}.
\end{align}
The expression for the applied load is then obtained from Eq.~(\ref{P-hinged}) together with Eq.~(\ref{eq-p0}):
\begin{equation}
\label{p-periodic}
P = q^2 +(1-2k^2) q^{-2}.
\end{equation}
From this infinity of possible solutions for a given value of $L$, only the one minimizing $P$ is selected, which fixes the value of $n$; see below. 

\begin{figure}
\includegraphics[width=\columnwidth]{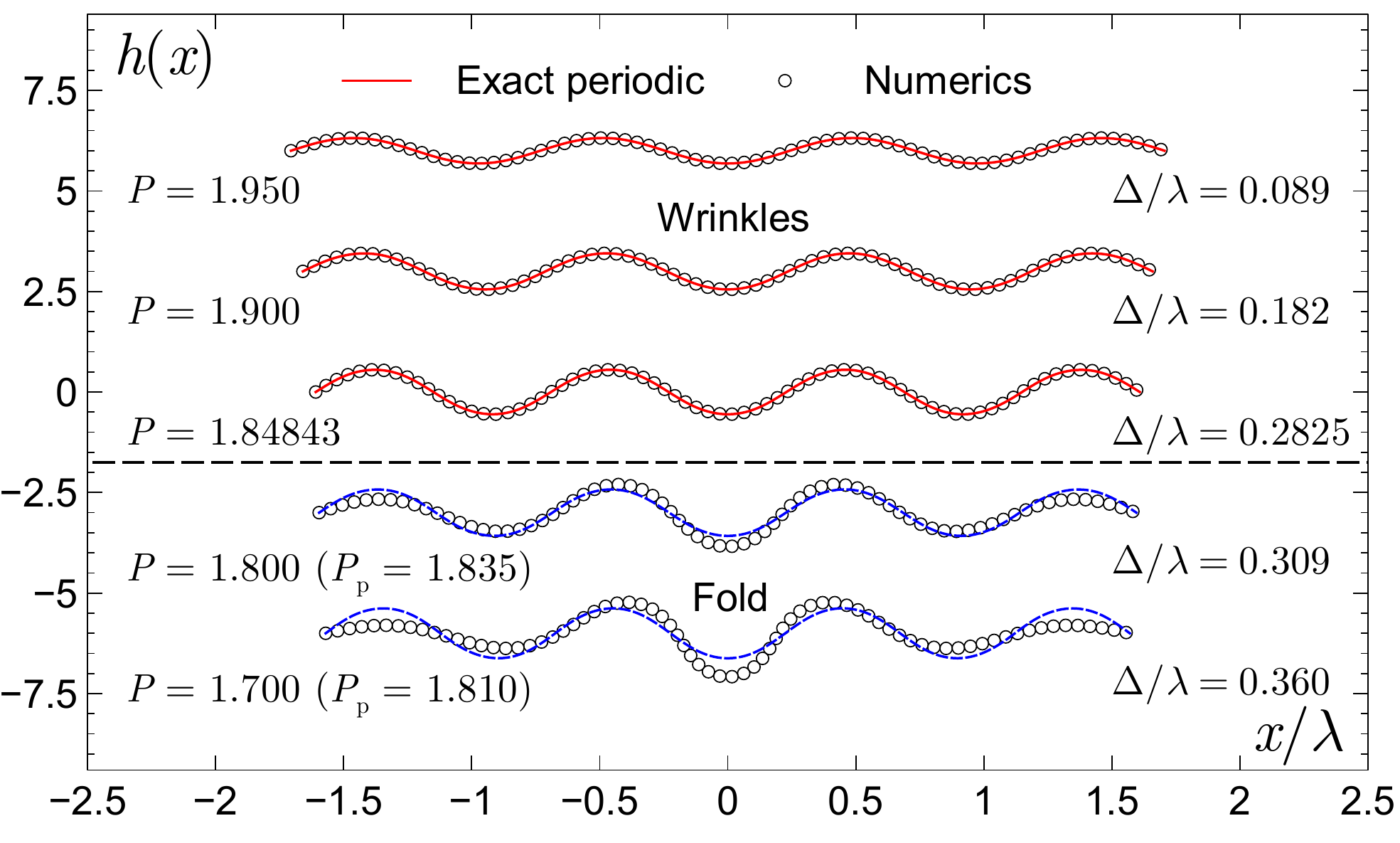}
\caption{(Color online) Evolution of the exact periodic solution for different values of $\Delta/\lambda$ and $L/\lambda = 3.5$ ($N=7$), where $x(s)$, $h(s)$ are given by Eqs.~(\ref{param-eq-x}) and (\ref{profile-period-hinged}) respectively and $\phi(s)$ is given by Eq.~(\ref{phi-profile-period-hinged}). The corresponding evolution for the numerical solution of Eq.~(\ref{main-equation}) is also shown. $P$ and $P_{\text{p}}$ correspond to the numerical and the periodic solutions respectively ($P=P_{\text{p}}$ for $\Delta/\lambda \le 0.2825$). The wrinkle-to-fold transition occurs at $\Delta/\lambda \simeq 0.2825$.}
\label{fig02}
\end{figure}

The parameter $k$ is related to the confinement $\Delta$ by using Eq.~(\ref{delta-period}):
\begin{equation}
\Delta = 2L -\frac{2}{q}\left[{\cal E}((n+ \varepsilon) K,k) + {\cal E}((n-\varepsilon) K,k)\right],
\end{equation}
where we used the expression (\ref{def-q}) for $q$, $q s_0=\varepsilon K(k)$, and the fact that ${\cal E}$ is an odd function.
%see Eq.~(\ref{jaco-eps-prop4})
When $n=2p+1$ such that $\varepsilon =0$, we obtain 
\begin{align}
\Delta &= 2L -\frac{4}{q}{\cal E}((2p+1) K,k)=2L -\frac{4}{q}(2p+1)E(k) \nonumber \\
&= 2L \left(1-\frac{E(k)}{K(k)}\right),
\end{align}
where $E(k)$ is the complete elliptic integral of the second kind~\cite[p.487]{olve10}. When $n=2p$ such that $\varepsilon =1$, we have
\begin{align}
\Delta &= 2L -\frac{2}{q}\left[{\cal E}((2p+1) K,k) + {\cal E}(2p-1) K,k)\right] \\
\label{delta-period-final}
&= 2L -\frac{8p}{q} E(k) = 2L \left(1-\frac{E(k)}{K(k)}\right).
\end{align}
Therefore, the relation between $k$ and $\Delta$ is the same for symmetric and antisymmetric solutions. The decrease of the applied pressure $P$ as a function of the confinement $\Delta$ is given by the parametric equation $(\Delta(k),P(k))$ given by Eqs.~(\ref{p-periodic}) and (\ref{delta-period-final}) and using Eq.~(\ref{def-q}). 

However, we still have to determine the optimal value of $n$ minimizing $P$. For $L=N \pi$ ($L/\lambda = N/2$), it can be shown that $n=N$ provided $k$ is small enough. Above some threshold, $k=k^{\star}$, we have $n=N-1$. As $\Delta(k)$ is an increasing function of $k$ [Eq.~(\ref{delta-period-final})] this means that for small confinement we have $n=N$ before reaching a threshold, $\Delta(k^{\star})$, above which the compressed system prefers to remove half a wavelength from the profile, $n=N-1$. However, if $\Delta(k^{\star})\ge \Delta_{\text{F}}$, where $\Delta_{\text{F}}$ is the critical confinement for which the periodic solution is unstable against the localized solution, then $n=N$ for all the values of confinement where the periodic solution is stable. The threshold $k^{\star}$ can be obtained by searching for which value of $k$, the pressure $P$ has the same value for $n=N$ and $n=N-1$. We found that $\Delta(k^{\star})/L \simeq (k^{\star})^2=2\pi/3L = \lambda/3L$, where we used a first order expansion for the confinement which is enough for this discussion (using the full expression leads to the same conclusion). Comparison with $\Delta_{\text{F}}/L= \lambda^2 /L^2$, which is derived in the next section, shows that for $L\ge 3 \lambda$ the optimal value of $n$ is always $n=N$ (with $L=N\pi$). Notice that it leads also to $k\le 1/3$. From now on, we assume that the length of the sheet is at least as large as three times the wrinkle wavelength. As a result, when $N$ is odd, the solution is symmetric and when $N$ is even, it is antisymmetric. For $L=N\pi$, we thus have $n=N$ and the expression for the pressure reads
\begin{equation}
\label{p-period-final}
P=\frac{4}{\pi^2} K(k)^2 + \frac{\pi^2 (1-2k^2)}{4 K(k)^2}.
\end{equation}

Let us summarize the scheme for calculating the exact periodic solution. Given $L=N\pi$, we have $n=N$. Given $\Delta/L$ we find $k$ form Eq.~(\ref{delta-period-final}). The values obtained for $n$ and $k$ are substituted in Eq.~(\ref{def-q}) to obtain $q$. These values of $k$ and $q$ are used in Eqs.~(\ref{profile-period-hinged}) and (\ref{p-periodic}) to obtain the height profile and the pressure. This solution is unique thanks to the monotonic-increasing nature of the right-hand-side of Eq.~(\ref{delta-period-final}).  Finally, the pressure-displacement relation is obtained parametrically from Eqs.~(\ref{delta-period-final}) and (\ref{p-period-final}); see Fig.~\ref{fig04}. Figure~\ref{fig02} shows the evolution with increasing confinement of the exact periodic solution for $L=7\pi$ ($L/\lambda = 3.5$) comparing it to the numerical solution of Eq.~(\ref{main-equation}). For $\Delta/\lambda \lesssim 0.2825$, the numerical solution is periodic. This threshold is close to the critical wrinkle-to-fold confinement computed in Section~\ref{sec_fold} for large sheets, $\Delta_{\text{F}}/\lambda = 0.2857$. 

Since it is expected that this exact periodic solution is unstable against localization for large enough confinement, we give below its expansion for small $\Delta$. We expand Eqs.~(\ref{delta-period-final}) and (\ref{p-period-final}) in small $k$ to obtain $\Delta=L k^2 +\mathcal{O}(k^4)$ and $P=2 -2k^2 + \mathcal{O}(k^4)$. Thus, to leading order in small relative confinement, we have
\begin{equation}
\label{pressure-wrinkle}
P \simeq 2\left(1-\frac{\Delta}{L}\right).
\end{equation}
The amplitude of the profile is $A=2k/q$ (see Eq.~(\ref{profile-period-hinged})). For $L=N\pi$, we have
\begin{equation}
A =\frac{\pi k}{K(k)}\simeq 2\sqrt{\frac{\Delta}{L}}.
\end{equation}
The corresponding profiles for small confinement are
\begin{subequations}
\begin{align}
\label{small-wrinkle-sym}
h(s) &= A \cos s, \quad \text{symmetric profile}, \\
\label{small-wrinkle-asym}
h(s) &= A \sin s, \quad \text{asymmetric profile}.
\end{align}
\end{subequations}
Thus, to leading order in $\Delta/L$, our exact solution reproduces the expressions for $P$, $A$, and $h(s)$, known for the wrinkle state \cite{miln89,huan10,zhang07,cerda04}. Notice that, because we consider $L\ge 3 \lambda$, and a transition to the localized solution occurs for $\Delta/L = \lambda^2 /L^2$, the maximum relative compression relevant for the periodic solution is $\Delta/L \le 1/9 \simeq 0.11$, which is indeed small. Therefore, a first-order expansion is quite sufficient for describing the wrinkle state.

\section{Approximate localized solutions}
%----------------------------------------
\label{sec_fold}

\subsection{Construction of the localized solution}

We have not been able to find exact physical solutions for folds in a finite sheet. Exact localized solutions of Eq.~(\ref{main-equation}) for a finite sheet do exist; yet, these profiles are not solutions of the present physical problem, as they have a finite height at the boundaries. We describe them in Appendix~\ref{sec_exact_solution} for the sake of mathematical interest and possible relevance for other systems to be studied in the future.

We resort to a perturbative calculation, extending the multiple-scale analysis of Ref.~\cite{audo11} to a finite system. The main motivation is to enable an accurate analysis of the wrinkle-to-fold transition, as will be presented in Sec.~\ref{sec_transition}. We therefore assume a very long (yet finite) sheet compared to the wrinkle wavelength, $L\gg \lambda$. It has been established that, as $L$ is made larger, the region of stable wrinkles shrinks and their amplitude diminishes \cite{arxiv10,audo11}. Thus, the critical values of the pressure and displacement for the wrinkle-to-fold transition, $P_{\text{F}}$ and $\Delta_{\text{F}}$, can be assumed arbitrarily close to the ones for the flat-to-wrinkle transition, $P_{\text{W}}=2$ and $\Delta_{\text{W}}=0$, as $L$ is taken to be arbitrarily large.

At the transition from a wrinkle state to a localized one, we anticipate that the undulations of wavelength $\lambda$ will be attenuated over a much larger length scale of order $L$. Therefore, the sheet profile contains two length scales, a short one ($\lambda$) and a long one ($L$). To obtain this profile near the transition, we then substitute in the energy, Eqs.~(\ref{ener-h})-(\ref{def-g}), a multiple-scale function of the form
\begin{equation}
\label{multiplescale}
 h(s) = \cos(k_{\text{c}} s) \sum_{j=1}^\infty \epsilon^j H_j(\epsilon s) \simeq \epsilon\, \cos(k_{\text{c}} s) H_1(\epsilon s),
\end{equation}
and consider hinged boundary conditions,
\begin{equation}
h(\pm L/2)=\ddot{h}(\pm L/2)=0.
\end{equation}
In this expansion, the small parameter $\epsilon$ is given by 
\begin{equation}
\label{def-eps}
  \epsilon=(P_{\text{W}}-P)^{1/2},
\end{equation}
where $P_{\text{W}}=2$ is the critical flat-to-wrinkle pressure. $k_c$
is the wavenumber of the fast oscillations; from the known
flat-to-wrinkle transition we expect to get $k_c=1$
\cite{miln89,poci08}. We have selected a symmetric profile; an
antisymmetric one is obtained by replacing the cosine with a sine and
leads to a similar analysis. For simplicity, we restrict the
discussion to commensurate sheets, $L=N\pi$ (\textit{i.e.}
$L/\lambda=N/2$), where odd and even $N$ correspond respectively to
symmetric and antisymmetric solutions. (We will later check the effect
of this approximation numerically.)

As in Ref.~\cite{audo11}, we substitute the profile
Eq.~(\ref{multiplescale}) in Eq.~(\ref{def-g}), expand in powers of
$\epsilon$, and integrate over the fast oscillations (which cancel the
terms proportional to $\epsilon$ and $\epsilon^3$), to obtain a
systematic expansion of the energy,
\begin{equation}
G=G_0 \epsilon^2 + G_1 \epsilon^4 + \mathcal{O}(\epsilon^6),
\end{equation}
Analysis of the leading order reproduces the known wrinkling
transition with $P_{\text{W}}=2$ and $k_{\text{c}}=1$ ($\lambda =
2\pi$) without any constraint on $H_1$. This gives $G_0=0$, reducing
the energy of the system to $G=G_1 \epsilon^4$. The
function $H_1$ is determined by minimizing the functional $G_1$, given
by
\begin{equation}
 G_1 \simeq \frac{1}{2}\int_{-L/2}^{L/2} \left( [H_1'(S)]^2 - \frac{1}{16} H_1^4 + \frac{1}{4} H_1^2 \right) ds,
\label{energy4}
\end{equation}
where $S\equiv\epsilon s$, and a prime denotes a derivative with respect to $S$. Variation of Eq.~(\ref{energy4}) gives the amplitude equation for $H_1$ \cite{audo11},
\begin{equation}
\label{amplitude}
 H_1''(S) + \frac{1}{8}H_1^3 - \frac{1}{4}H_1 = 0.
\end{equation}
The boundary conditions of vanishing height, $h(\pm L/2)=0$, are automatically satisfied since $\cos((N/2)\pi)=0$ for odd $N$. The boundary conditions of vanishing bending moment, $\ddot{h}(\pm L/2)=0$, impose
\begin{equation}
 H_1'(S=\pm\epsilon L/2) = 0.
\label{bcH}
\end{equation}

Equations (\ref{amplitude}) and (\ref{bcH}) always have the trivial constant solution, $H_1=\sqrt{2}$, corresponding to wrinkles. In addition, Eq.~(\ref{amplitude}) has solutions in terms of Jacobian elliptic functions~\cite{Abramowitz}. Out of the twelve Jacobian functions, only one is found to provide a physical solution,\footnote{The Jacobian function dc yields a solution identical to Eq.~(\ref{amplitudesolution}). The function cn describes a solution
whose amplitude {\it decreases} with increasing displacement. The remaining nine functions either diverge with increasing $L$ and/or are odd, thus reversing the symmetry of the solution.}
\begin{equation}
\label{amplitudesolution} 
 H_1(S) = 4\kappa\, \text{dn} (\kappa S,m), \quad \kappa = \frac{1}{2\sqrt{2-m}}.
\end{equation}
In this section using the other definition of the modulus $m$~\cite{Abramowitz} makes the presentation more elegant. It is related to the modulus $k$ of Sec.~\ref{sec_wrinkle} by $m=k^2$. The modulus $m$ ($0\leq m\leq 1$) changes with the displacement $\Delta$ (see Eq.~(\ref{Deltafold}) below). The pressure is also related to the modulus $m$ using the boundary condition (\ref{bcH}),\footnote{We use here the envelope at its lowest mode (no nodes). Higher modes, satisfying $\kappa \epsilon L = 2nK(m)$ with $n>1$, are valid solutions but with higher energy.}
\begin{equation}
\label{bcm}
 \kappa \epsilon L = 2K(m),
\end{equation}
where $K(m)$ is the complete elliptic integral of the first kind~\cite{Abramowitz}, which is half the period of the function $\text{dn}$~\cite{Abramowitz}, $\kappa(m)$ is defined in Eq.~(\ref{amplitudesolution}), and $\epsilon = \sqrt{2-P}$. Equations (\ref{multiplescale}), (\ref{amplitudesolution}) and (\ref{bcm}) yield the height profile, $h(s)$, of the sheet for a given $m$,
\begin{subequations}
\begin{align}
\label{localized-profile-multiscale}
h(s)&= 4\kappa \epsilon\,   \text{dn}(\kappa \epsilon s,m)\, \cos (s), \\
\label{localized-profile}
&= \frac{8K(m)}{L} \, \text{dn}\left(\frac{2K(m)}{L} s,m\right) \, \cos (s),
\end{align}
\end{subequations}
for $L/\lambda = N/2$.

\begin{figure}
\includegraphics[width=\columnwidth]{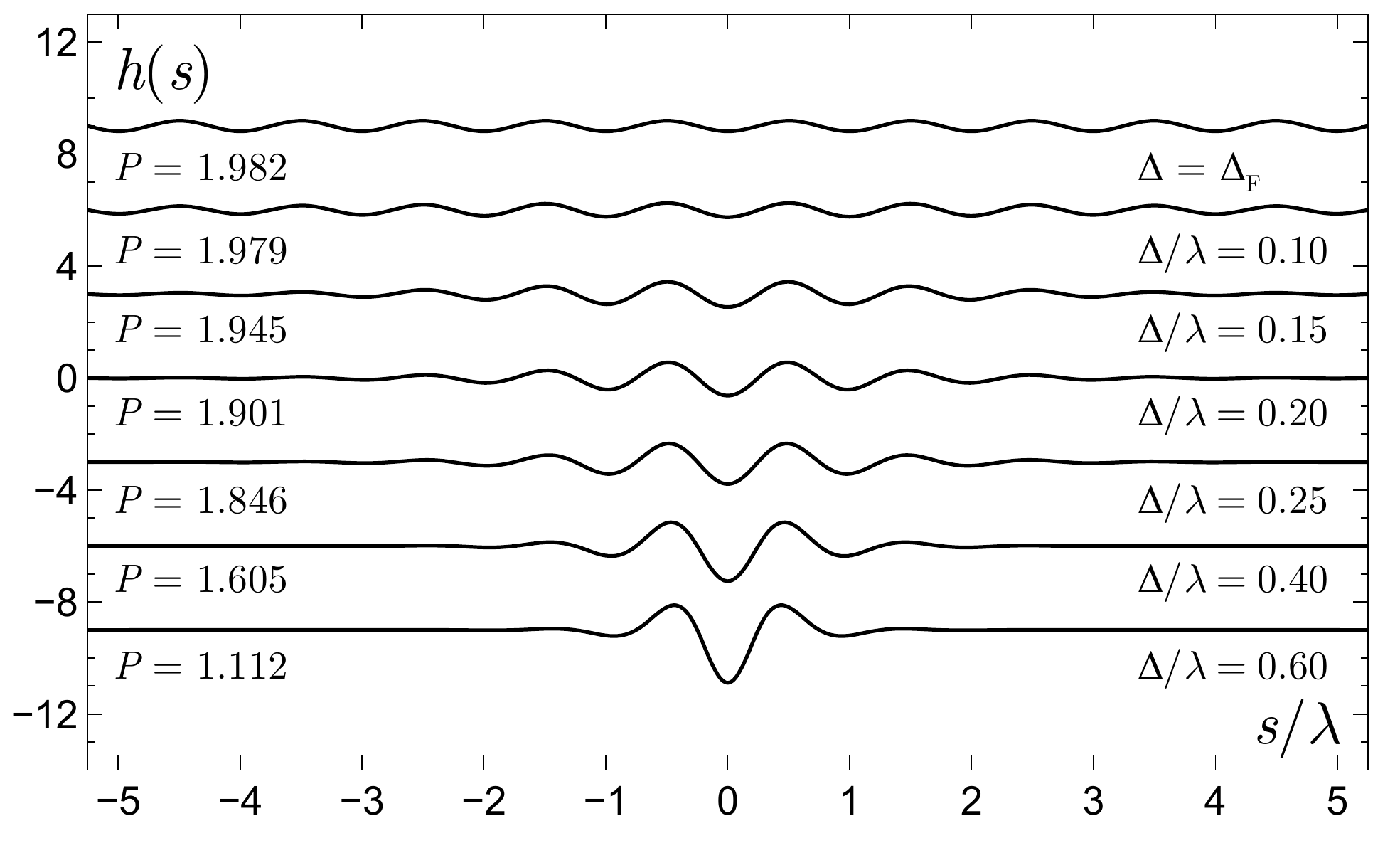}
\caption{Localized height profiles for a sheet of length $L/\lambda=10.5$ for various values of the confinement parameter.}
\label{fig03}
\end{figure}

\begin{figure}
\includegraphics[width=\columnwidth]{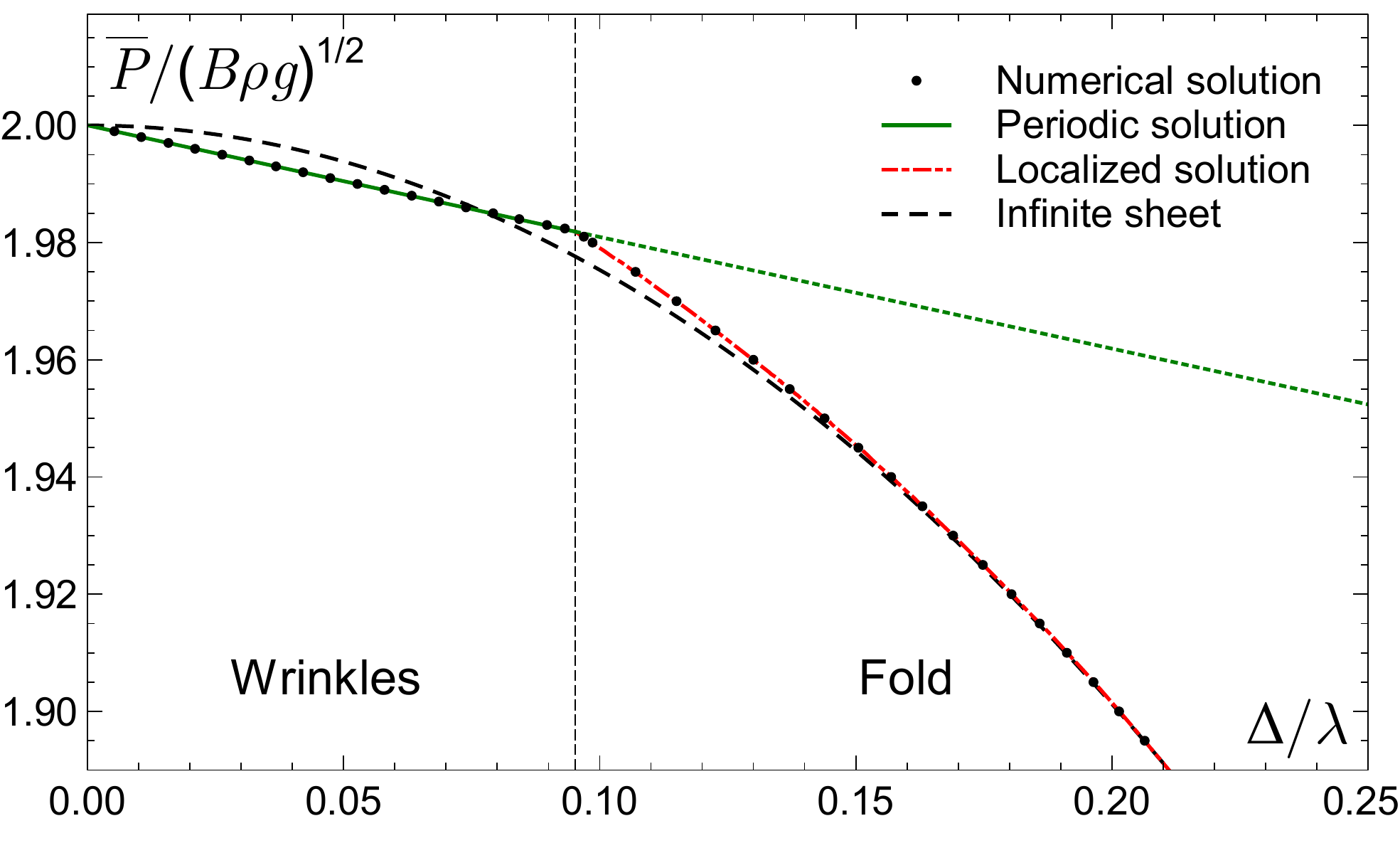}
\caption{(Color online) Pressure as a function of displacement for a sheet of length $L/\lambda=10.5$, showing the second-order transition between the wrinkle state (exact solution, green solid line) and the fold state (multiple scale approximation, red dash-dotted line). $\bar{P}$ is the physical dimensional pressure. Numerical solution (circles) of Eq.~(\ref{main-equation}) indicates that the parametric equations (\ref{delta-period-final}) and (\ref{p-period-final}) for the wrinkle state, and (\ref{pressure-fold}) and (\ref{Deltafold}) for the fold state, provide the correct evolution of $P$ as a function of $\Delta$. The pressure-displacement relation for a confined infinite sheet, Eq.~(\ref{p-infinite}), is shown for comparison (exact solution, black dashed line).}
\label{fig04}
\end{figure}

The resulting profiles are demonstrated in Fig.~\ref{fig03}. In the limit $m\rightarrow 0$, the function $\text{dn}(u,m)\rightarrow 1$, and from Eq.~(\ref{localized-profile}) we recover the wrinkled profile, $h(s)=\sqrt{2}\epsilon\cos (s)$ \cite{arxiv10}. In the opposite limit, $m\rightarrow 1$, we have $\text{dn}(u,m)\rightarrow 1/\cosh(u)$, which recovers the approximate localized fold for an infinite sheet \cite{arxiv10,audo11}, $h(s)=2\epsilon\cos(s)/\cosh(\epsilon s/2)$. These two limits are further treated in Sec.~\ref{sec_pressure_fold}. Thus, $m$ serves as the order parameter of the wrinkle-to-fold transition, as will be discussed in Sec.~\ref{sec_transition}. The antisymmetric counterpart of Eq.~(\ref{localized-profile}) is obtained by replacing $\cos (s)$ with $\sin (s)$. 

To further characterize the extent of localization, we define the decay length of the envelope as,
\begin{equation}
\xi \equiv \left|\frac{H_1(0)}{\ddot{H}_1(0)}\right|^{1/2}=\frac{L}{2\sqrt{m}K(m)},
\label{localization-length}
\end{equation}
where the last equality follows from
Eq.~(\ref{localized-profile}). The decay length
  diverges ($\xi\rightarrow \infty$) when $m\rightarrow 0$, and
  vanishes ($\xi \rightarrow 0$) as $m\rightarrow 1$. This defines
the two limits of weakly and strongly localized profiles to be
discussed below. Note, however, that the leading order of the
multiple-scale analysis is not strictly valid in the strongly
localized limit. The crossover between these two limits, $\xi=L$,
where the localization should become observable, occurs at $m\simeq
0.096$.

\subsection{Pressure, displacement, and amplitude}
\label{sec_pressure_fold}

Using Eq.~(\ref{bcm}) together with Eq.~(\ref{def-eps}), we obtain the
expression for the pressure associated with the localized solution,
\begin{equation}
\label{pressure-fold}
P(m)=2-\frac{16}{L^2}(2-m) K(m)^2.
\end{equation} 
Since we know that for an infinite sheet the wrinkle state is always unstable against the localized state~\cite{audo11}, we expect that this transition should occur at an arbitrarily small displacement for a sufficiently long sheet. Therefore, we expand the expression (\ref{delta-h}) of the displacement to the leading order in $\epsilon$,
\begin{equation} 
\label{delta-first-order}
 \Delta \simeq \frac{1}{2} \int_{-L/2}^{L/2} ds\, \dot{h}^2.
\end{equation}
Using the expression (\ref{localized-profile-multiscale}) for $h$, averaging over the fast oscillations (short length scale), and performing the integral over the slow envelope (long length scale), we obtain
\begin{equation}
\label{Deltam}
 \Delta = 8\kappa \epsilon\, {\cal E}(\kappa\epsilon L/2,m),
\end{equation}
where ${\cal E}(x,m)$ is the Jacobi epsilon function, and $\kappa$ is given as a function of $m$ in Eq.~(\ref{amplitudesolution}). Using Eq.~(\ref{bcm}), the displacement associated with the localized solution reads,
\begin{equation}
\label{Deltafold}
 \Delta(m) = 8\kappa \epsilon\, {\cal E}(K(m),m) = \frac{16}{L} K(m) E(m),
\end{equation}
where $E(m)$ is the complete elliptic integral of the second kind~\cite{Abramowitz}. Since both $\Delta$, given by Eq.~(\ref{Deltafold}), and $P$, given by Eq.~(\ref{pressure-fold}), are functions of $m$, the evolution of the pressure as a function of the displacement is  given by the parametric form ($\Delta(m)$, $P(m)$). This evolution is demonstrated in Fig.~\ref{fig04}.

\subsubsection{Weakly localized limit}
\label{sec_limit_m0}

As $m$ grows from zero the profile ceases to be purely periodic and begins to localize. In the weakly localized limit, $m\ll 1$, the symmetry has already been broken, but the localization length is still larger than the system size, $\xi > L$. In this limit, therefore, the deviation from the wrinkle state will be hard to resolve experimentally. Nevertheless, as mentioned above, $m\simeq 0.1$ is sufficient to reach $\xi \simeq L$.

The displacement given by Eq.~(\ref{Deltafold}) is an increasing function of $m$. We notice that, in the limit $m \to 0$, the displacement takes a finite value,
\begin{equation}
\label{critical-disp}
\Delta_{\text{F}}\equiv \Delta(0)= 4\pi^2 /L = \lambda^2/L.
\end{equation}
This means that the localized solution emerges only beyond this finite
displacement in the case of sheets of finite length. For a confinement
smaller than $\Delta_{\text{F}}$, the shape of the sheet is described
by the periodic solution constructed in Sec.~\ref{sec_wrinkle}. In
Appendix \ref{scaling} we obtain the critical wrinkle-to-fold
confinement, up to a constant prefactor, $\Delta_{\text{F}} \sim
\lambda^2/L$, from a simple scaling analysis.

For small $m$ we have
\begin{align}
\label{expansion-delta-loc}
\Delta(m) &= \Delta_{\text{F}} + \frac{\pi^2 m^2}{8L} + \mathcal{O}(m^3), \\
\label{expansion-p-loc}
P(m) &= 2-\frac{8\pi^2}{L^2}-\frac{3\pi^2 m^2}{4 L^2} + \mathcal{O}(m^3).
\end{align}
The parameter $m$ can be eliminated to obtain
\begin{equation}
P \simeq 2 - \frac{2\Delta_{\text{F}}}{L}-6 \left(\frac{\Delta-\Delta_{\text{F}}}{L} \right).
\end{equation}
The profile for $m=0$ is obtained from Eq.~(\ref{localized-profile}),
\begin{equation}
h(s) = \frac{4\pi}{L} \cos s, \quad m=0.
\end{equation}
Although this profile describes periodic wrinkles, it has a finite amplitude since it is obtained for a finite displacement, $\Delta=\Delta_{\text{F}}$. 

Once $m>0$, the profile is no longer periodic and its extrema have different heights. The height of the central fold, $A_0 = |h(0)|$, and that of the extremum next to it, $A_1 \simeq |h(\pi)|$, are of special interest since they are used to experimentally characterize the evolution of the localized profile~\cite{poci08}; see also Sec.~\ref{sec_discuss} and Fig.~\ref{fig07}. (The localization makes the second extremum shift slightly from $s=\pi$, but this shift is of higher order than our present approximation.) From Eqs.~(\ref{localized-profile}) and (\ref{expansion-delta-loc}) we get 
%%\begin{subequations}
%%\label{A0-A1-m}
%%\begin{eqnarray}
%%A_0 &=& \frac{4\pi}{L}+\frac{\pi m}{L}+ \mathcal{O}(m^2), \\
%%A_1 &=& \frac{4\pi}{L}+\cos\left(\frac{2\pi^2}{L}\right)\frac{\pi m}{L}+ \mathcal{O}(m^2).
%%\end{eqnarray}
%%\end{subequations}
%%Eliminating the parameter $m$ between Eq.~(\ref{expansion-delta-loc}) and the expansion of the amplitudes we obtain
\begin{subequations}
\label{A0-A1-delta}
\begin{align}
A_0 &\simeq \frac{4\pi}{L}+2\sqrt{2} \left(\frac{\Delta - \Delta_{\text{F}}}{L}\right)^{1/2}, \\
A_1 &\simeq \frac{4\pi}{L}+2\sqrt{2}\cos\left(\frac{2\pi^2}{L}\right)\left(\frac{\Delta - \Delta_{\text{F}}}{L}\right)^{1/2},
\end{align}
\end{subequations}
which are valid for $\Delta \gtrsim \Delta_{\text{F}}$. The slopes of both $A_0$ and $A_1$ diverge at $\Delta_{\text{F}}$, with an exponent of $1/2$, as is typical for a mean-field second-order transition.

\subsubsection{Strongly localized limit}

In this limit of $m \rightarrow 1$ we have
\begin{subequations}
\begin{align}
\Delta(m) &\simeq \frac{16}{L} K(m), \\
P(m) &\simeq 2-\frac{16}{L^2} K(m)^2.
\end{align}
\end{subequations}
Eliminating the parameter $m$ between these two last relations leads to
\begin{equation}
\label{p-infinite}
P=2-\frac{\Delta^2}{16}=2-\frac{\pi^2}{4}\left(\frac{\Delta}{\lambda}\right)^2,
\end{equation}
which coincides with the exact pressure-displacement relation for infinite sheets~\cite{diam11}. In this limit, we also have $\text{dn}(u,m)\rightarrow 1/\cosh(u)$~\cite{Abramowitz}, such that the profile~(\ref{localized-profile}) becomes
\begin{equation}
h(s)=\frac{\Delta}{2} \frac{\cos (s)}{\cosh(s/\xi)}, \quad \xi=8/\Delta, 
\end{equation}
which coincides with the approximate localized fold for an infinite sheet~\cite{arxiv10,audo11}. The amplitude $A_0$ is thus given by the following simple relation,
\begin{equation}
A_0 = \Delta/2,
\label{A_0m1}
\end{equation}
which again is identical to the exact result for an infinite sheet \cite{diam11}.

\section{Wrinkle-to-fold transition}
%-----------------------------------
\label{sec_transition}

Below a critical value of $\epsilon L=\epsilon_{\text{F}}
L=2^{3/2}\pi$, Eq.\ (\ref{bcm}) has no solution for $m$. Thus, below
this critical confinement, the localized solution does not exist, and
the only possible envelope is the trivial constant one, corresponding
to wrinkles. At $\epsilon_{\text{F}} L$ itself Eq.\ (\ref{bcm}) admits
$m=0$ as a unique solution, and for larger values of $\epsilon L$,
with $m>0$ as a unique solution, the envelope grows continuously,
corresponding to increasingly localized patterns. Accordingly, we
define the order parameter of the wrinkle-to-fold transition as $m$,
and its control parameter as
 \begin{equation}
\label{control-parameter}
\tau\equiv\epsilon L, \quad \tau_{\text{F}}=\epsilon_{\text{F}} L = 2^{3/2}\pi.
\end{equation}

\begin{figure*}
\includegraphics[width=\textwidth]{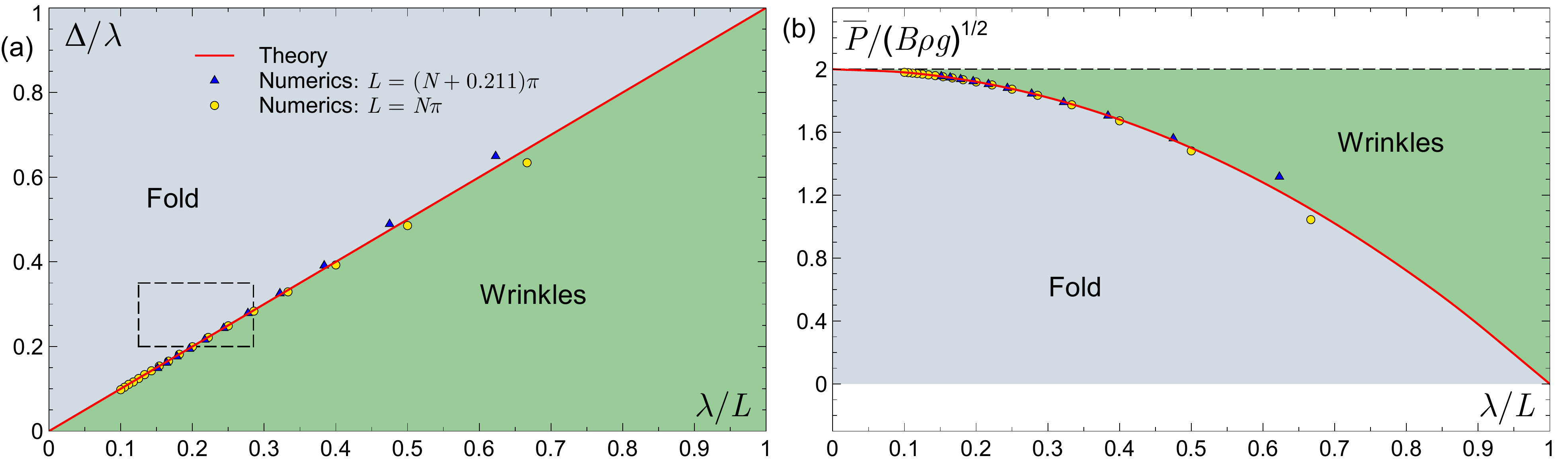}
\caption{(Color online) Phase diagrams of an incompressible
  fluid-supported sheet. (a) Phase diagram on the displacement--length
  plane. For $0<\Delta<\lambda^2/L$ the sheet forms extended wrinkles,
  and for $\Delta>\lambda^2 /L$ a localized fold is formed. The
  rectangle shows the range of the experimentally determined phase
  boundary \cite{poci08} (see also Fig.~\ref{fig07}). (b) Phase
  diagram on the pressure--length plane. $\bar{P}$ is the physical
  dimensional pressure. Wrinkles are stable for
  $2-2\lambda^2/L^2<P<2$, and a fold is stable for $P<2-2\lambda^2
  /L^2$. The wrinkle-fold line is a second-order transition. Note how
  the region of stable wrinkles in both diagrams vanishes in the limit
  of an infinite sheet. In both diagrams, the theoretical transition
  curves are compared to the numerical transition obtained by solving
  Eq.~(\ref{main-equation}) and analyzing the periodicity of the
  numerical solution from the relative height of its
  extrema. Numerical results for both commensurate ($L$ being an integer
  or half-integer multiple of $\lambda$) and incommensurate sheets are
  presented, showing a negligible discrepancy between these two and
  the theoretical prediction for $L\gtrsim 2\lambda$.}
\label{fig05}
\end{figure*}

\subsection{Landau expansion}
\label{sec_landau}

A clear way to present the continuous transition is through a Landau
expansion of the energy, $G$, in small order parameter. The Landau
expansion is to be performed on the energy {\it prior to
  minimization}. For this purpose, we use the following form for the
envelope:
\begin{equation}
\label{H1star}
H_1^{\star}(S) = \frac{2}{\sqrt{2-m}}\, \text{dn} \left(\frac{2K(m)}{\tau} S,m\right),
\end{equation}
where we have substituted the boundary condition (\ref{bcm}) in the
argument of the profile given by Eq.~(\ref{amplitudesolution}), but
not in its amplitude. Therefore, $H_1^{\star}(S)$ satisfies the
boundary conditions but does not yet minimize the energy.

Substituting Eq.~(\ref{H1star}) in Eq.~(\ref{energy4}) and expanding
to fourth order in small $m$, yields
\begin{equation}
\label{Landau}
 \Delta G \simeq (\tau_{\text{F}}^2-\tau^2)(m^2 + m^3) + \frac{3}{128}(35\tau_{\text{F}}^2-33\tau^2)m^4,
\end{equation}
where $\Delta G= 32L(G-G_{\text{W}})/\epsilon^2$, and $G_{\text{W}}=L\epsilon^4/4$ is the energy of the wrinkles.
 The appearance of the $m^3$ term has no significance for the transition
because of its vanishing prefactor. Upon minimization of $\Delta G$ with respect to $m$ we obtain,
\begin{equation}
 m \simeq 8 \left[ (\tau-\tau_{\text{F}})/3\tau_{\text{F}} \right]^{1/2},
\label{mtau}
\end{equation}
with a critical exponent, $\beta=1/2$, as usual for a mean-field second-order
transition.

Using the expression (\ref{control-parameter}) of $\tau_{\text{F}}$ in
Eqs.~(\ref{pressure-wrinkle}) and (\ref{def-eps}) for $P$ and
$\Delta$, we recover the critical values derived in
Sec.~\ref{sec_limit_m0} from the analysis of weakly localized
folds. We repeat here these central results:
\begin{align}
\label{Pcc}
 P_{\text{F}} &= P_{\text{W}} - \tau_{\text{F}}^2/L^2 = 2 - 8\pi^2/L^2, \\
\label{Deltacc}
 \Delta_{\text{F}} &= \tau_{\text{F}}^2/(2L) = 4\pi^2/L.
\end{align}
The value of $ \Delta_{\text{F}} \simeq 39.5/L$ corrects an earlier, higher estimate of $\Delta_{\text{F}} = 48/L$, which was obtained by using an Ansatz profile that was not an energy minimizer \cite{arxiv10}. Equations (\ref{Pcc}) and (\ref{Deltacc}) confirm that the wrinkle-to-fold transition can be made arbitrarily close to the flat-to-wrinkle one with increasing size of the sheet. Consequently, this separate second-order transition appears only in finite systems. 

In Fig.~\ref{fig05} we summarize the phase diagrams of the incompressible fluid-supported sheet on the $\Delta$--$L$ and $P$--$L$ planes. Note that these scaled diagrams are parameter-free. 

\begin{figure*}
\includegraphics[width=\textwidth]{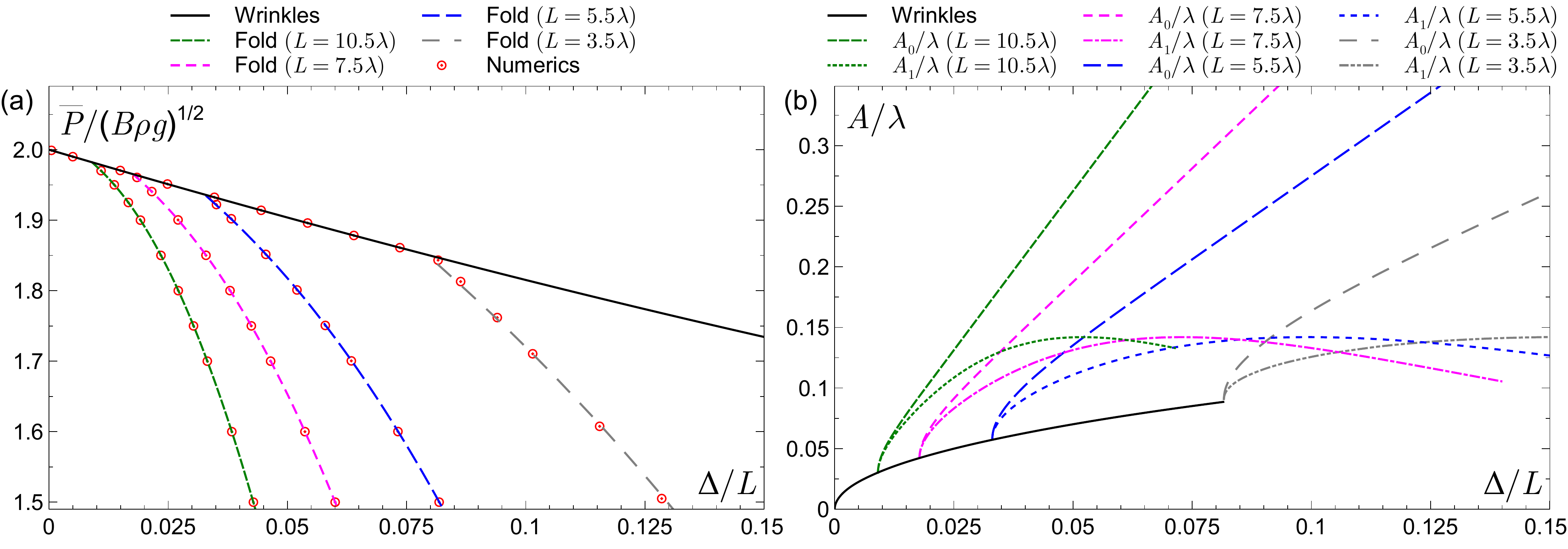}
\caption{(Color online) (a) Evolution of the rescaled pressure as a function of the relative displacement for various sheet lengths $L$. $\bar{P}$ is the physical dimensional pressure. The graphs show the evolution of the pressure for the periodic state (wrinkles) together with the bifurcations toward the localized state (fold) at $\Delta/L=\Delta_{\text{F}}/L=\lambda^2/L^2$ and $P=2-2\Delta_{\text{F}}/L$. These bifurcations decrease the energy of the system compared to the periodic state. (b) Evolution of the amplitudes of the pattern, rescaled by the wrinkles wavelength, as a function of the relative displacement for the same sheet lengths as those used in panel (a). The wrinkles amplitude grows as $(\Delta/L)^{1/2}$ until the points $\Delta/L=\lambda^2/L^2$ and $A/\lambda= \lambda/L\pi$ is reached, where the localized state emerges. The evolution of the amplitude of the central fold, $A_0$, and of the fold next to it, $A_1$, are shown.}
\label{fig06}
\end{figure*}

\subsection{Matching at the transition}
\label{sec_matching}

Below the wrinkle-to-fold transition and considering small wrinkles (to first order in $\Delta/L$), we have found in Sec.~\ref{sec_pressure_period}, 
\begin{equation}
\label{P-small-wrinkles}
P = 2\left(1-\frac{\Delta}{L}\right), \quad A_0 = 2\sqrt{\frac{\Delta}{L}}, \quad h(s) = A_0 \cos (s).
\end{equation}
At the wrinkle-to-fold-transition, we have shown that $\Delta=\Delta_{\text{F}}=\lambda^2/L$ and $P=P _{\text{F}}= 2-2\lambda^2/L^2$. Slightly above the transition, the weakly localized solution is characterized by (see Sec.~\ref{sec_limit_m0}),
\begin{align}
\label{P-above-transition}
P &= 2\left(1-\frac{\Delta_{\text{F}}}{L}\right)-6 \left(\frac{\Delta-\Delta_{\text{F}}}{L} \right), \\
A_0 &= \frac{2\lambda}{L}+2\sqrt{2} \left(\frac{\Delta - \Delta_{\text{F}}}{L}\right)^{1/2}, \nonumber\\
h(s) &= \frac{2\lambda}{L} \cos (s), \quad \text{at the transition} \nonumber
\end{align}
Comparing Eqs.~(\ref{P-small-wrinkles}) and (\ref{P-above-transition}), we see that the transition is continuous. The discontinuity appears in the pressure derivative, with $(dP/d\Delta)_{\Delta\rightarrow\Delta_{\text{F}}^-}=-2/L$ and $(dP/d\Delta)_{\Delta\rightarrow\Delta_{\text{F}}^+}=-6/L$. Hence, the continuous transition is of second order (discontinuity in the second derivative of $E$ with respect to $\Delta$). 
The pressure associated with the localized solution is smaller than the one associated with the periodic solution once $\Delta> \Delta_{\text{F}}$, as illustrated in Fig~\ref{fig06}a. Since $P=dE/d\Delta$, the localized state has a smaller energy compared to the energy of the periodic state. The latter is thus unstable once $\Delta > \Delta_{\text{F}}$. The evolution of the amplitudes of the wrinkle and localized states as the displacement increases is shown in Fig~\ref{fig06}b.

%{\color{blue}{Moved to discussion - The theory developed here is thus correct at the first order in $\Delta/L$. A higher order theory would not change the main conclusions we obtained. It would perhaps slightly modify some quantitative predictions like the value of the confinement, $\Delta_{\text{F}}$, at which the transition occurs. However, we doubt that the difference between this theory and a higher order one could be easily discriminated experimentally. Confrontation of this first order theory with accurate experimental data will indicate if the theory is precise enough.}}

\section{Discussion}
%-------------------
\label{sec_discuss}
\subsection{Experimental consequences}

%\footnoteRemainder: The value of $m$ in which the localization length becomes comparable with the system size can be approximated as follows. Define the localization length by, $\tilde{\kappa}=-\left[\frac{\ddot{h}(0)}{h(0)}\right]^{1/2}$, and using Eq.~(\ref{localized-profile}), gives $\tilde{\kappa}^2=\frac{4mK(m)^2}{L^2}$. Experimentally, the localized state is observed when $\tilde{\kappa}\lesssim \frac{1}{L}$. Solving this inequality for $m$ we obtain $m^{\star}\gtrsim 0.1$. Then, by Eq.~(\ref{pressure-fold}), the observed critical pressure is $P_{\text{F}}(m^{\star}\simeq 0.1\div 0.3)\gtrsim 2-(1.001\div 1.011)\frac{8\pi^2}{L}$, and the observed critical displacement: $\Delta_{\text{F}}(m^{\star}\simeq 0.1\div 0.3)\gtrsim (1.0035\div 1.0039)\frac{4\pi^2}{L}$. Very close to the theoretical values.

The system parameters are the length $L$ and bending modulus $B$ of the thin elastic sheet, and the density $\rho$ of the liquid. The experimental control parameter is either the displacement $\Delta$ or the force per unit length $P$. The measured quantity is the height profile, including the wrinkles wavelength, $\lambda$, and height extrema ~\cite{poci08}. 

The order parameter $m$, which we have defined here, characterizes the decay of the envelope away from the center of the fold. Experimentally, one may look instead at the height difference between two consecutive local extrema $A_0$ and $A_1$~\cite{poci08}. These two definitions of the order parameter are equivalent close to the transition. Indeed, from Eqs.~(\ref{expansion-delta-loc}) and (\ref{A0-A1-delta}) we find
\begin{align}
 M_{\text{exp}} &\equiv 1 - \frac{A_1}{A_0} = \frac{1}{2} [\sin(\lambda \pi/2L)]^2 m +\mathcal{O}(m^2), \nonumber \\
 &= \frac{\pi^2 \lambda^2}{8 L^2}m + \mathcal{O}(m^2,\lambda^4/L^4).
\end{align}
Using Eq.~(\ref{expansion-delta-loc}), or equivalently Eq.~(\ref{mtau}), this result is written as a function of the experimental control parameter $\Delta$:
\begin{equation}
M_{\text{exp}} = \frac{\pi^2 }{\sqrt{2}} \frac{\lambda}{L}\left(\frac{\Delta-\Delta_{\text{F}}}{L}\right)^{1/2},
\end{equation}
where $\Delta_{\text{F}}$ is given by Eq.~(\ref{critical-disp}).

\begin{figure}
\includegraphics[width=\columnwidth]{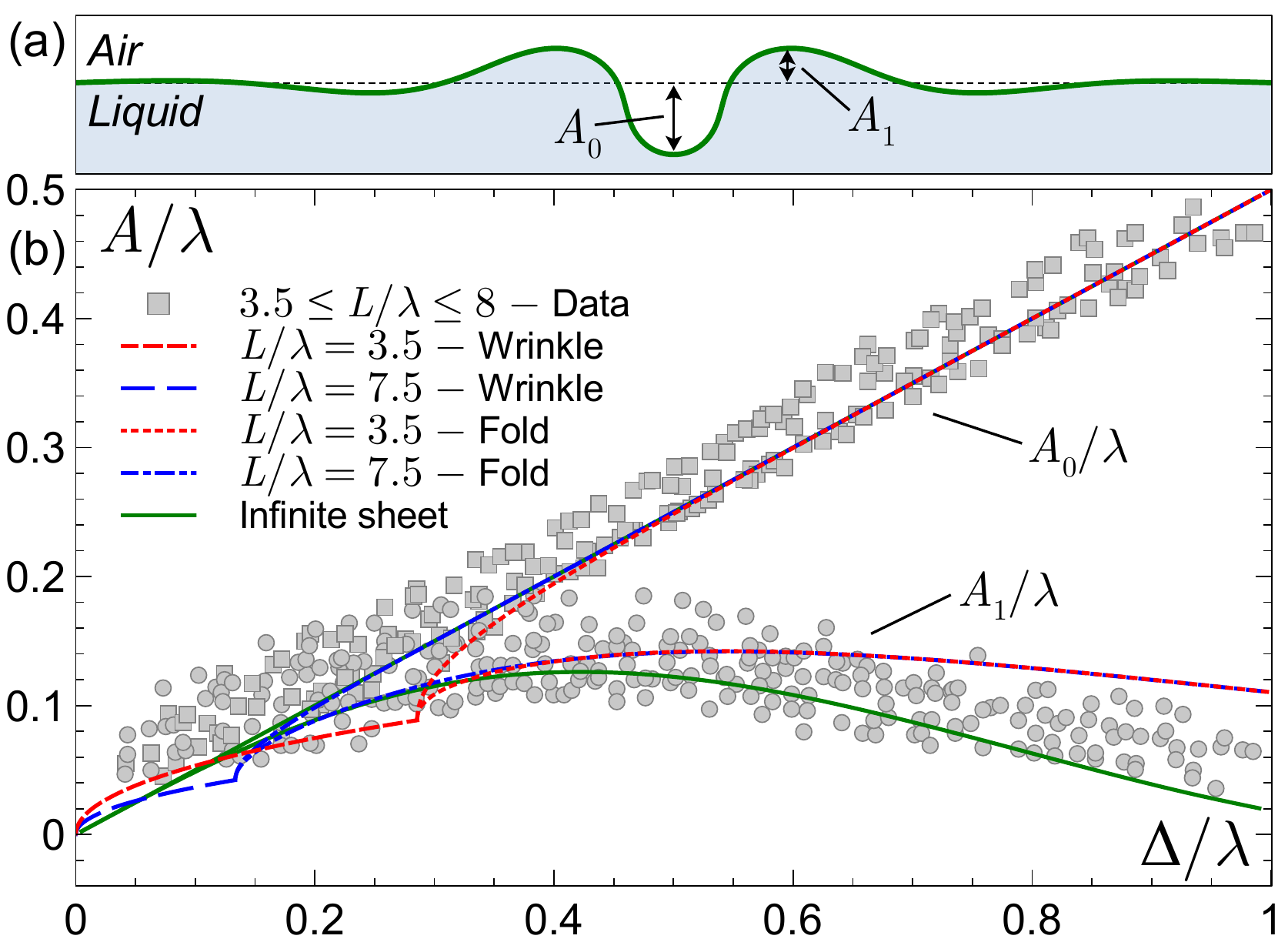}
\caption{(Color online) (a) Definitions of the amplitudes $A_0$ and $A_1$ for the localized state. (b) Comparison between the experimental evolution of $A_0$ and $A_1$ (rescaled by $\lambda$) with increasing confinement for finite sheets~\cite{poci08}, and the prediction of Eq.~(\ref{localized-profile}) where $A_0=|h(0)|$ and $A_1=|h(\pi)|$. The evolution of the amplitudes of the exact solution for an infinite sheet is also presented~\cite{diam11}.}
\label{fig07}
\end{figure}

The evolution of the pattern amplitude as a function of the displacement can also be measured. Such data are available in Ref.~\cite{poci08}, whose main purpose was to show the universality of the folding state where all the data collapsed onto a master curve for large enough confinement. Indeed, in the large-confinement regime, the sheet deformation is localized in a small region, comparable to the wrinkle wavelength $\lambda$, and does not depend significantly on the sheet length $L$, provided that $L$ is sufficiently large compared to $\lambda$. The data of Ref.~\cite{poci08} are reproduced in Fig.~\ref{fig07} together with the evolution of the amplitudes predicted by the present theory. (Note that, to comply with the presentation of the experimental data, the amplitude in Fig.~\ref{fig07} is drawn as a function of $\Delta/\lambda$ rather than $\Delta/L$ as in Fig.~\ref{fig06}b.) 

Our theory is valid for large enough sheets and consequently for a sufficiently small relative confinement $\Delta/L \ll 1$. See, for example, the deviation of the amplitude $A_1$ at large $\Delta/L$ for a small system with $L/\lambda=3.5$ (Fig.~\ref{fig07}). On the other hand, at large $\Delta/\lambda$, as the localization length becomes significantly smaller than $L$, the exact predictions for an infinite sheet~\cite{diam11} become accurate for finite sheets as well (see Fig.~\ref{fig07}). Finally, for the localized state we have assumed a commensurate sheet ($L$ being an integer or half-integer multiple of $\lambda$). Yet, as shown in Fig.~\ref{fig05}, the effect of incommensurability becomes negligible already for sheets larger than $2\lambda$, making it inconsequential experimentally.  

%While it should describe accurately the second order transition between periodic and localized states, it should not describe the pattern for high confinement except the evolution of $A_0$ which appears to coincide with the evolution of an infinitely long sheet. As seen in Fig.~\ref{fig07}, accurate data focusing on the region where the transition is predicted to occur are needed to validate the present theory ($\Delta$ ranging from 0 to a value slightly above $\Delta_{\text{F}}$). The predicted evolution of the amplitudes as a function $\Delta/L$, such that the growth of the amplitude of periodic state does not depend on the sheet length (see Fig.~\ref{fig06}), reads

The authors of Ref.~\cite{poci08} inferred from their experiments that the wrinkle-to-fold transition occurred at a value of $\Delta\simeq 0.3\lambda$ for all sheet lengths $L$, whereas our theory gives an $L$-dependent critical displacement (see Fig.~\ref{fig05}a).    
As seen in Fig.~\ref{fig07}, it was natural to draw that conclusion given the experimental error. The rectangle shown in  the diagram of Fig.~\ref{fig05}a, representing the range of the experimentally determined thresholds (Fig.~\ref{fig07}), is consistent with the theoretical prediction. Clearly, accurate experimental data focusing on the transition region are still needed. For this purpose we summarize the predictions for the amplitude evolution below and above the transition: 
\begin{subequations}
\begin{align}
\text{For}\ &\Delta < \Delta_{\text{F}}: \nonumber \\
\frac{A}{\lambda} &= \frac{1}{\pi}\left(\frac{\Delta}{L}\right)^{1/2}, \\
\text{For}\ &\Delta \gtrsim \Delta_{\text{F}}: \nonumber \\
\frac{A_0}{\lambda} &= \frac{\lambda}{\pi L}+\frac{\sqrt{2}}{\pi} \left(\frac{\Delta - \Delta_{\text{F}}}{L}\right)^{1/2}, \\
\frac{A_1}{\lambda} &= \frac{\lambda}{\pi L}+\frac{\sqrt{2}}{\pi} \cos\left(\frac{2\pi^2}{L}\right)\left(\frac{\Delta - \Delta_{\text{F}}}{L}\right)^{1/2}.
\end{align}
\end{subequations}

In Sec.~\ref{sec_fold}, we derived the properties of the critical wrinkle-to-fold transition at which the localization length $\xi$ diverges. As for critical phenomena in a finite system, the transition will be observable in practice only when the localization length becomes smaller than the system size, $\xi\lesssim L$. We find, however, that the displacement required to get $\xi\simeq L$ is larger than $\Delta_{\text{F}}$ by only $0.4\%$. Thus, the finite size of the system should have a negligible effect on the experimentally observed critical displacement.

\subsection{Conclusion}

In the present work we have derived an exact solution for the wrinkle state in a finite fluid-supported sheet. The theory developed for fold localization, however, is restricted to large sheets; it is correct to first order in $\lambda/L$. A higher-order theory (next order in $\epsilon$ of the multiple-scale expansion) would not change the main conclusions obtained above. It is expected to slightly modify some quantitative predictions such as the value of the critical wrinkle-to-fold confinement, $\Delta_{\text{F}}$. The possibility to discriminate experimentally between the present theory of the transition and a higher-order one is doubtful. 

Although we found exact localized solutions for the governing equation (\ref{main-equation}), these do not give the physical profile of the sheet (see Appendix~\ref{sec_exact_solution}). Mathematically, the inability to derive physical localized profiles from these solutions may indicate that such profiles belong to a different branch of solutions to the non-linear governing equation. Such a branch is yet to be found. If it exists, it will probably coincide, in the limit $L\rightarrow \infty$, with the exact solution known for this limit \cite{diam11}.

An important extension to the present theory is the inclusion of finite compressibility of the sheet. This will allow us to treat the flat-to-wrinkle transition, and thus construct the complete ``phase-diagram'' of the floating sheet including the two transitions. This extension will be presented in a forthcoming publication.

\begin{acknowledgments}
We thank Philip Rosenau for a helpful discussion. This work has been supported in part by the Israel Science Foundation (Grant No.\ 164/14).
\end{acknowledgments}

\appendix 

\section{Exact localized solutions to Eq.~(\ref{main-equation})} 
%---------------------------------------------------------------
\label{sec_exact_solution}
In this Appendix we derive exact localized solutions to Eq.~(\ref{main-equation}), which are not physical solutions of our problem treated here, but may be of use in future research. 
Equation~(\ref{main-equation}) is a member of the sine-Gordon-modified-Korteweg-de Vries hierarchy~\cite{gesz03} in which the sine-Gordon equation is a lower member. Consequently, a solution of the sine-Gordon equation is also a solution of Eq.~(\ref{main-equation}). Using the separation of variable proposed by Lamb~\cite{lamb71}, the following function
\begin{equation}
\label{ansatz}
\phi(s,t) = 4 \arctan(F(s)/G(t)),
\end{equation}
is a solution of the sine-Gordon equation provided $F$ and $G$ satisfy the following differential equations~\cite{dele80}
\begin{subequations}
\label{f-and-g}
\begin{align}
(F')^2 &= -\kappa F^4 + \mu F^2+\lambda, \\
(G')^2 &= \kappa G^4 + (\mu-1) G^2 - \lambda,
\end{align}
\end{subequations}
where $\kappa$, $\mu$ and $\lambda$ are arbitrary constants. The function (\ref{ansatz}) with $t=s$ is also a solution of Eq.~(\ref{main-equation}) if $F$ and $G$ satisfy the relations (\ref{f-and-g}) and provided that
\begin{equation}
\label{P-localized}
\tilde{P} = 2- 4\mu.
\end{equation}
Introducing some scaling parameters~\cite{cost78}, $A f(\beta s)=F(s)$ and $g(\omega s) = 1/G(s)$, Eqs.~(\ref{f-and-g}) become
\begin{subequations}
\label{f-and-g2}
\begin{align}
(f')^2 &= \beta^{-2}\left[-\kappa A^2 f^4 + \mu f^2+\lambda A^{-2}\right], \\
(g')^2 &= \omega^{-2}\left[- \lambda g^4+ (\mu-1) g^2+ \kappa  \right].
\end{align}
\end{subequations}
Solving Eqs.~(\ref{f-and-g2}) allows to determine $f$ and $g$ and to obtain an exact solution of Eq.~(\ref{main-equation}). Those equations are satisfied by the Jacobi elliptic functions with arguments $\beta s$ and $\omega s$ and parameters $k_f$ and $k_g$. The various parameters are fixed by comparison between Eqs.~(\ref{f-and-g2}) and those satisfied by the Jacobi functions:
\begin{subequations}
\label{jacobi}
\begin{align}
\label{sn}
\text{sn}(s,k): \ (y')^2 &= k^2 y^4 -(1+k^2)y^2 +1 \\
\label{cn}
\text{cn}(s,k): \ (y')^2 &= -k^2 y^4 +(2k^2-1)y^2 +1-k^2 \\
\label{dn}
\text{dn}(s,k): \ (y')^2 &= - y^4 +(2-k^2)y^2 -1+k^2
\end{align}
\end{subequations}

\subsection*{Hinged sheets}

\begin{figure}[!t]
\includegraphics[width=\columnwidth]{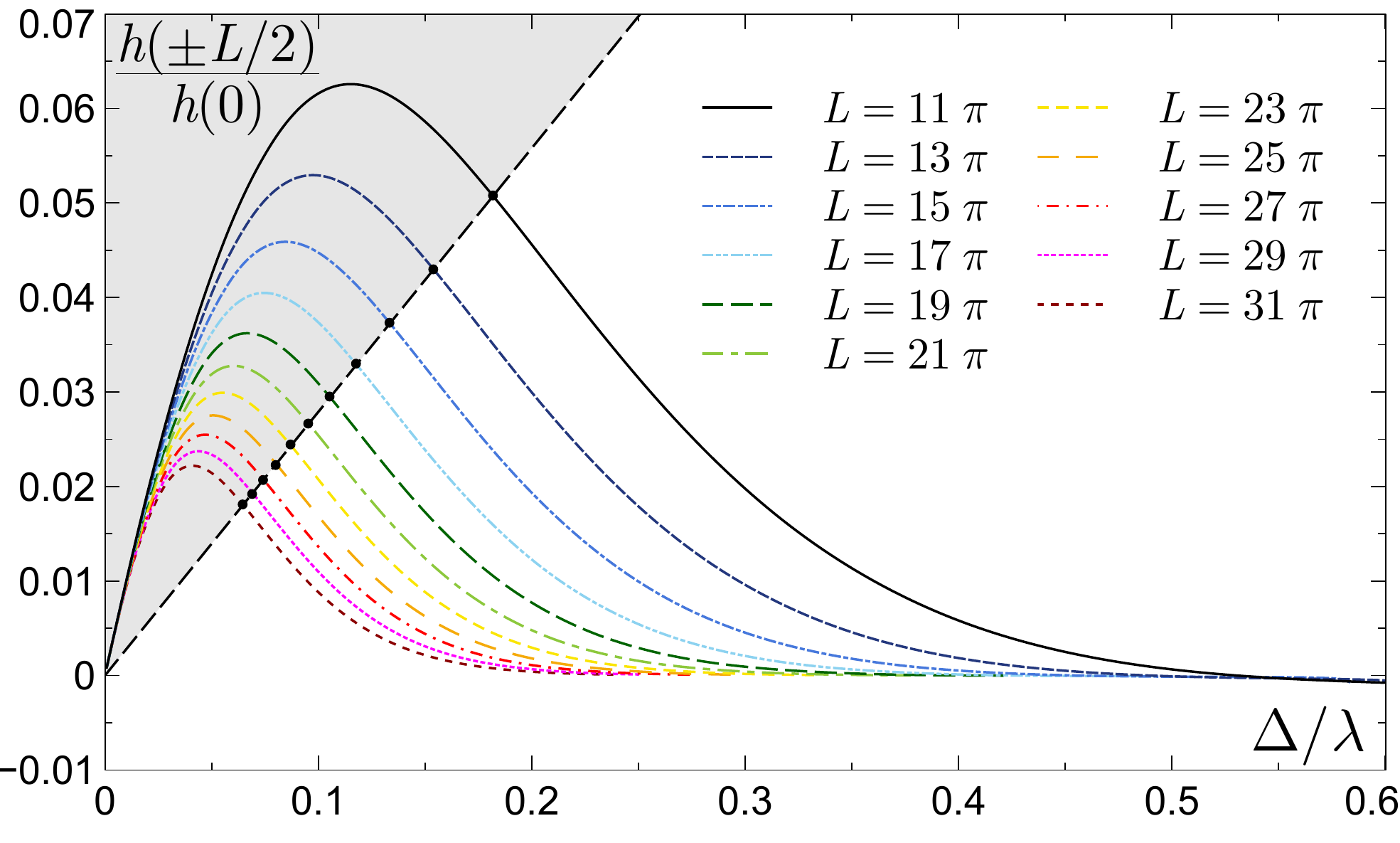}
\caption{(Color online) Evolution of the normalized height of the exact solution (\ref{sol-hinged-localized-sym}) at the boundaries as a function of the normalized displacement. The dots indicate the position of the threshold, $\Delta_{\text{F}}/\lambda$, below which the profile is periodic. The shaded area corresponds thus to the wrinkle state. This area is delimited by the relation $h(\pm L/2)/h(0) = 0.279\, \Delta_{\text{F}}/\lambda = 0.279\, \lambda/L$.}
\label{fig08}
\end{figure}

We choose $f=\text{cn}(\beta s, k_f)$ and $g=\text{sn}(\omega^{+} s,k^{+}_g)$ or $g=\text{cn}(\omega^{-} s,k^{-}_g)$ such that
\begin{subequations}
\label{sol-hinged-localized}
\begin{align}
\label{sol-hinged-localized-sym}
\phi(s) &= 4 \arctan[A\, \text{cn}(\beta s, k_f)\, \text{sn}(\omega^{+} s,k^{+}_g)], \\
\label{sol-hinged-localized-asym}
\phi(s) &= 4 \arctan[A\, \text{cn}(\beta s, k_f)\, \text{cn}(\omega^{-} s,k^{-}_g)].
\end{align}
\end{subequations}
Equations (\ref{sol-hinged-localized-sym}) and (\ref{sol-hinged-localized-asym}) lead to symmetric and antisymmetric solutions respectively. Comparison between Eqs.~(\ref{f-and-g2}) and Eqs.~(\ref{jacobi}) for that particular choices of $f$ and $g$ fixes six parameters (among eight parameters for each parity of $\phi$) as a function of $A$ and $\beta$ as follows:
\begin{subequations}
\label{parameters-hinged}
\begin{align}
\label{om-hinged-sym}
\omega^{+} &=\frac{\left[1 + (1 + A^2) \beta^2 \right]^{1/2}}{1 + A^2}, \\
\label{om-hinged-asym}
\omega^{-} &=\left[\frac{2}{1 + A^2} + \beta^2 -1 \right]^{1/2}, \\
\label{kg-hinged-sym}
k_g^{+} &= A \left[\frac{1 + A^2}{1 + (1 + A^2) \beta^2} -1 \right]^{1/2}, \\
\label{kg-hinged-asym}
k_g^{-} &= A \left[\frac{\beta^2+A^2(\beta^2-1)}{1 -A^4+ (1 + A^2)^2 \beta^2}\right]^{1/2}, \\
\label{kf-hinged}
k_f &= \frac{A}{1 + A^2} \left(1 + A^2 + \beta^{-2} \right)^{1/2}, \\
\label{mu-hinged}
\mu &= \frac{2 A^2 + (A^4-1) \beta^2}{(1 + A^2)^2}.
\end{align}
\end{subequations}
The expressions of $\kappa$ and $\lambda$ are not needed in the following and are not written. The amplitude $A$ of the solutions (\ref{sol-hinged-localized}) is related to the displacement $\Delta$ through Eq.~(\ref{delta-phi}) whereas $\beta$ should be fixed to satisfy the boundary conditions.

\begin{figure}[!t]
\includegraphics[width=\columnwidth]{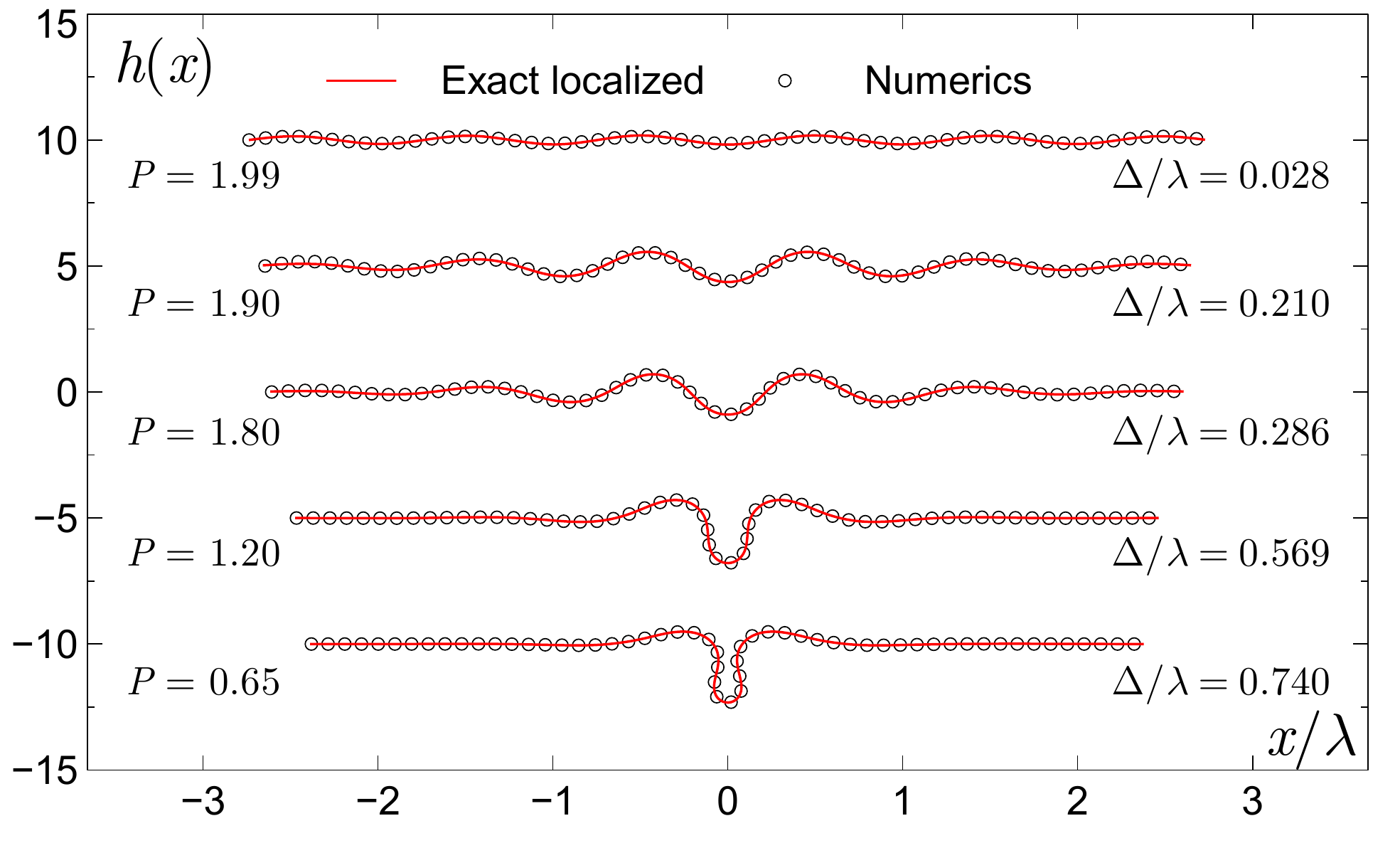}
\caption{(Color online) Evolution of the solution (\ref{sol-hinged-localized-sym}) for different values of $\Delta/\lambda$ and $L/\lambda = 5.5$, where $x(s)$, $h(s)$ are given by Eqs.~(\ref{param-eq-xh}) and $\phi(s)$ is given by Eq.~(\ref{sol-hinged-localized-sym}). The corresponding evolution for the numerical solution of Eq.~(\ref{main-equation}) is also shown. For $\Delta/\lambda < \Delta_{\text{F}}/\lambda = 0.182$, the numerical solution is periodic and coincides with the solution reported in Section~\ref{sec_wrinkle}.}
\label{fig09}
\end{figure}

For hinged boundary conditions both $h$ and $\ddot{h}=\dot{\phi} \cos \phi$ vanish at $s=\pm L/2$. From Eq.~(\ref{elastica-hydro}), this is equivalent to $\dot{\phi}(\pm L/2) = \dddot{\phi}(\pm L/2)=0$. The condition $\dot{\phi}(L/2)=0$ for the symmetric solution (\ref{sol-hinged-localized-sym}) leads to the constraint
\begin{equation}
\label{beta-eq}
\frac{\text{dn}(x,k_f)\, \text{sn}(x,k_f)\, \text{sn}(y,k^{+}_g)}{\text{cn}(x,k_f)\, \text{cn}(y,k^{+}_g)\, \text{dn}(y,k^{+}_g)}= \frac{y}{x},
\end{equation}
where $x= \beta L/2$ and $y=\omega^{+} L/2$. Due to the definite parity of the solution, $\dot{\phi}(-L/2)=0$ is automatically satisfied. A similar relation is easily obtained for the antisymmetric solution (\ref{sol-hinged-localized-asym}) which is not written here. For a given amplitude, $A$, and a given sheet length, $L$, Eq.~(\ref{beta-eq}) fixes $\beta$. The displacement, $\Delta$, and the pressure, $\tilde{P}$, are then computed from Eqs.~(\ref{delta-phi}) and (\ref{P-localized}) respectively. Finally, $P$ is computed from Eq.~(\ref{P-hinged}).

It is however impossible to satisfy both boundary conditions with the solutions (\ref{sol-hinged-localized}). For the exact periodic solution (\ref{sol-pendule}), both $\dot{\phi}(s)$ and $\dddot{\phi}(s)$ are proportional to $\text{cn}(q(s+s_0),k)$ such that they can simultaneously vanish at the boundaries with a suitable choice of $q$ (and $s_0$), see Eqs.~(\ref{q-s0}). Here, $\dot{\phi}(L/2)$ and $\dddot{\phi}(L/2)$ cannot simultaneously vanish for the same value of $\beta$. Therefore, the height of the profile assumes a finite value at the boundaries as shown in Fig.~\ref{fig08}. Notice however that $h(\pm L/2)/h(0)$ decreases as $L$ increases.

Nevertheless, despite the fact that it does not properly satisfy the boundary conditions, this solution is a very good approximation of the numerical solution satisfying both boundary conditions. Figure~\ref{fig09} shows a comparison between this exact localized solution and the numerical solution of Eq.~(\ref{main-equation}). The agreement is good especially near the central fold and for large enough confinement.

%Artificial constraint of ${\cal H}=0$.

\section{Accuracy of $\Delta_{\text{F}}$ and $P_{\text{F}}$} 
%---------------------------------------------------------------
\label{sec_acc_df_pf}

\begin{figure}[!t]
\includegraphics[width=\columnwidth]{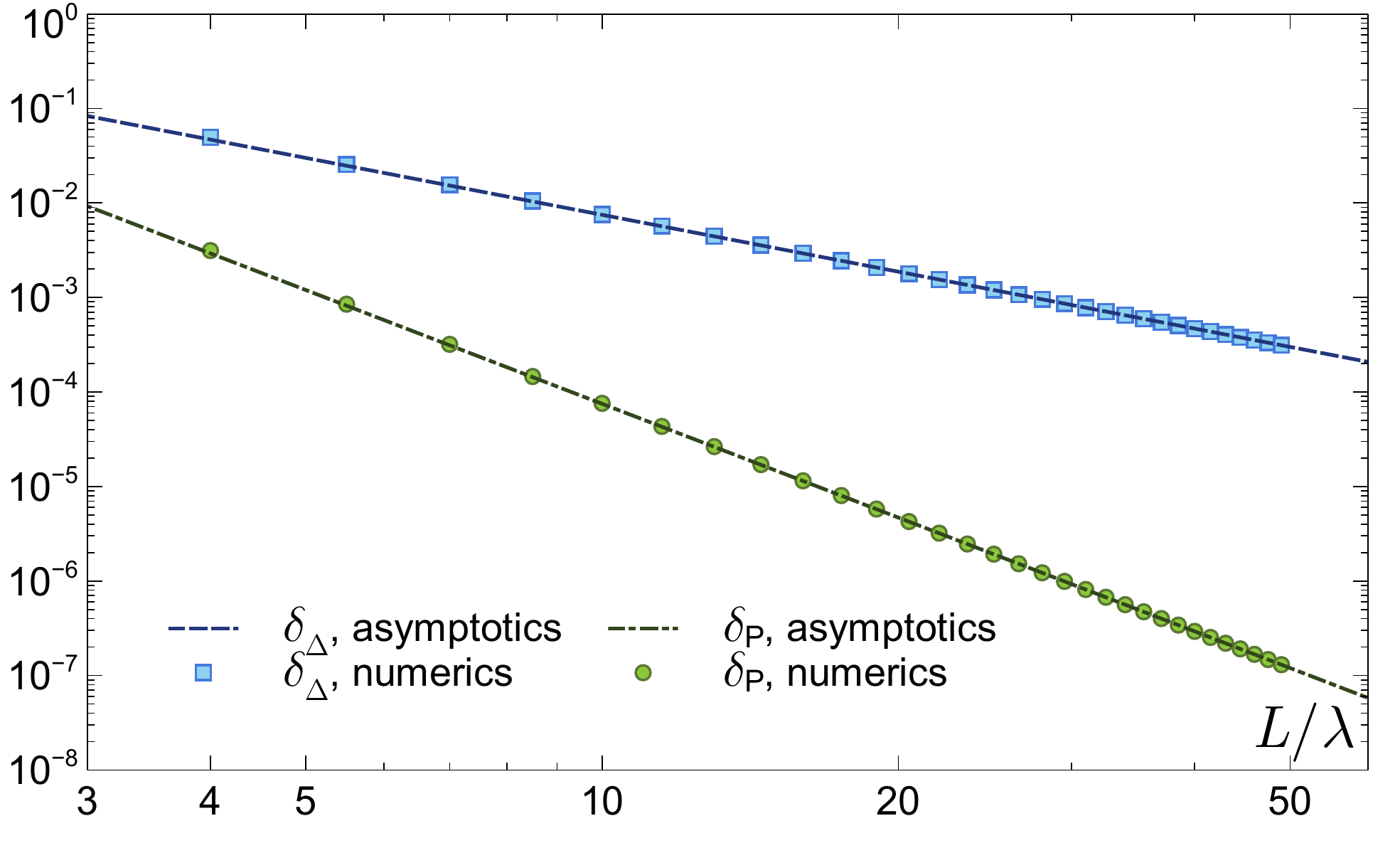}
\caption{(Color online) Evolution of $\delta_{\Delta}$ and $\delta_{\text{P}}$ as a function of the sheet length (normalized by the wavelength) obtained by solving numerically Eqs.~(\ref{def-delta}). The asymptotic behaviors (\ref{deltaDP-asymp}) are also shown.}
\label{fig10}
\end{figure}

The critical displacement, $\Delta_{\text{F}}$, and pressure, $P_{\text{F}}$, at which the transition from wrinkles to fold occurs has been computed in Sections \ref{sec_limit_m0} and \ref{sec_transition}. These expressions are obtained from matching, at the first order in $\Delta/L$, between the exact periodic and the approximate localized solutions, see Section~\ref{sec_matching}, and converge toward the exact values in the limit $L\to \infty$. Although a rigorous derivation of the error on these two quantities is only possible if their exact expressions are known, an estimation of the rate of convergence can be obtained from the exact periodic solution. 

Indeed, since at the transition $P(\Delta)$ coincides for both the periodic and the localized solutions, the exact expressions of $\Delta_{\text{F}}$ and $P_{\text{F}}$ should satisfy  Eqs.~(\ref{delta-period-final}) and (\ref{p-period-final}) for the same $k=k_{\text{c}}$. This is not the case with the approximate expressions we have obtained in this paper. Therefore, we define two quantities, $\delta_{\Delta}$ and $\delta_{\text{P}}$, as follows
\begin{subequations}
\label{def-delta}
\begin{align}
\label{def-deltaD}
\delta_{\Delta} &= \frac{\Delta(\bar{k}_{\text{c}}) - \Delta_{\text{F}}}{\Delta(\bar{k}_{\text{c}})} \quad \text{with} \quad P(\bar{k}_{\text{c}})= P_{\text{F}}, \\
\label{def-deltaP}
\delta_{\text{P}} &= \frac{P(\tilde{k}_{\text{c}}) - P_{\text{F}}}{P(\tilde{k}_{\text{c}})} \quad \text{with} \quad \Delta(\tilde{k}_{\text{c}})= \Delta_{\text{F}},
\end{align}
\end{subequations}
where $\Delta(k)$, $P(k)$, $P_{\text{F}}$ and $\Delta_{\text{F}}$ are given by Eqs.~(\ref{delta-period-final}), (\ref{p-period-final}), (\ref{Pcc}) and (\ref{Deltacc}) respectively. With the exact expressions of $\Delta_{\text{F}}$ and $P_{\text{F}}$, we have $\bar{k}_{\text{c}}=\tilde{k}_{\text{c}}$ and $\delta_{\Delta}=\delta_{\text{P}}=0$. The quantities $\delta_{\Delta}$ and $\delta_{\text{P}}$ are thus a measure of the error introduced and of the rate of convergence toward the exact expressions of the critical displacement and pressure. 

Expanding both $\Delta(k)$ and $P(k)$ up to $k^4$, we obtain at the leading order in $\lambda/L$:
\begin{equation}
\label{deltaDP-asymp}
\delta_{\Delta} =\frac{3\lambda^2}{4L^2} + \mathcal{O}\left[\frac{\lambda^4}{L^4}\right] \quad \text{and} \quad \delta_{\text{P}} = \frac{3\lambda^4}{4L^4} + \mathcal{O}\left[\frac{\lambda^6}{L^6}\right].
\end{equation}
Figure \ref{fig10} shows that the evolution of $\delta_{\Delta}$ and $\delta_{\text{P}}$ as a function of $L/\lambda$ obtained by solving numerically Eqs.~(\ref{def-delta}) agrees well with the asymptotic expression (\ref{deltaDP-asymp}) even when $L/\lambda$ is not so large. Therefore, the expressions of $\Delta_{\text{F}}$ and $P_{\text{F}}$ derived in the paper converges toward the exact values as $L^{-2}$ and $L^{-4}$ respectively.

\section{Scaling approach}
\label{scaling}

In this section, we show that some of the results obtained in the main text, and in particular the critical confinement at which a wrinkle-to-fold transition occurs, can be qualitatively recovered using a simple scaling approach. As formulated in Sec.~\ref{sec_system}, the bending energy of the sheet $E_{\text{b}}$ and the deformation energy of the substrate $E_{\text{s}}$ read
\begin{align}
E_{\text{b}} &= \frac{WB}{2} \int_{-L/2}^{L/2} \dot{\phi}^2 ds, \\
E_{\text{s}} &= \frac{W\rho g}{2} \int_{-L/2}^{L/2} h^2 \cos \phi\, ds,
\end{align}
and the displacement along the direction of confinement is given by
\begin{equation}
\Delta = \int_{-L/2}^{L/2} (1-\cos \phi)\, ds.
\end{equation}

\subsection{Small displacement}

\begin{figure}[!t]
\includegraphics[width=\columnwidth]{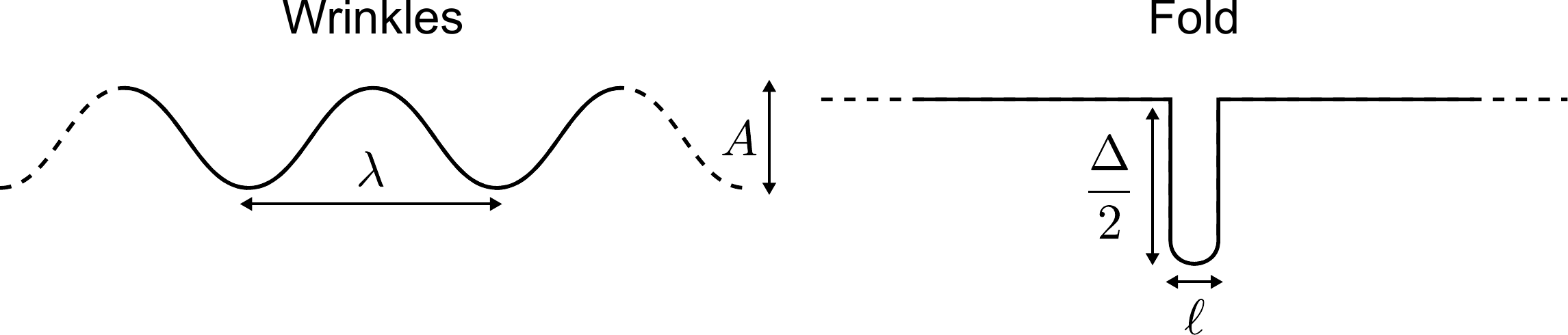}
\caption{Schematics for wrinkles and fold patterns.}
\label{fig11}
\end{figure}

For small displacement, leading to small sheet deformations ($\phi \ll 1$), wrinkles emerge with an amplitude $A$, wavelength $\lambda$ and curvature $\dot{\phi}\sim A/\lambda^2$, see Fig.~\ref{fig11}. The energies thus scale as 
\begin{equation}
E_{\text{b}} \sim BW L A^2 /\lambda^4; \quad E_{\text{s}} \sim \rho g W L A^2.
\end{equation}
The balance of these two energies leads to
\begin{equation}
\lambda \sim (B/ \rho g)^{1/4}.
\end{equation}
In the limit $\phi \ll 1$, the relation between the amplitude and the displacement is given by
\begin{equation}
\label{amp-wrinkle}
\Delta \sim \int_{-L/2}^{L/2} \phi^2 \, ds \sim \int_{-L/2}^{L/2} (dh/ds)^2 \, ds \sim L (A/\lambda)^2.
\end{equation}

\subsection{Large displacement}

The total energy of the system can be written as
\begin{align}
\frac{E}{W} &= \frac{B}{2} \int_{-L/2}^{L/2} \dot{\phi}^2 ds + \frac{\rho g}{2} \int_{-L/2}^{L/2} h^2 ds \\
&-\frac{\rho g}{2} \int_{-L/2}^{L/2} h^2 (1-\cos \phi\,) ds \\
&\sim B \Lambda \dot{\phi}^2 + \rho g \Lambda A^2 - \rho g A^2 \Delta, \\
&\sim \rho g \left[\lambda^4 \Lambda \dot{\phi}^2 + \Lambda A^2 - A^2 \Delta\right],
\end{align}
where $\Lambda$ is the spatial extent of the localized pattern, $\dot{\phi}$ its typical curvature and $A$ its amplitude. 

For wrinkles, we have $\Lambda \sim L$, $\dot{\phi} \sim A/\lambda^2$ and $A \sim \lambda \sqrt{\Delta/L}$, see Eq.~(\ref{amp-wrinkle}). The wrinkles energy can thus be written as
\begin{equation}
E_{W}/W = \rho g\lambda^3 \left[a\, \Delta/\lambda - b\, (\lambda/L) (\Delta/\lambda)^2\right],
\end{equation}
where $a$ and $b$ are numerical constants. For a fold, we have $\Lambda \sim \ell$, the fold width, $\dot{\phi} \sim \ell^{-1}$, and $A \sim \Delta$, see Fig.~\ref{fig11}. The fold energy can thus be written as
\begin{equation}
E_{F}/W = \rho g \left[(c^2/4\bar{c})\lambda^4/\ell + \bar{c}\, \ell \Delta^2- d \Delta^3\right],
\end{equation}
where $c$, $\bar{c}$ and $d$ are numerical constants. A minimization of $E_{F}$ with respect to $\ell$ gives $\ell = (c/2\bar{c}) \lambda^2 /\Delta$ and the energy becomes
\begin{equation}
E_{F}/W = \rho g\lambda^3 \left[c\, \Delta/\lambda- d (\Delta/\lambda)^3\right].
\end{equation}
Requiring that the wrinkles and fold energies match at $\Delta =
\Delta_{\text{F}}$ gives a quadratic equation for
$\Delta_{\text{F}}$. Demanding that $\Delta_{\text{F}}$ vanishes when
$L$ diverges yields $a=c$. A wrinkle-to-fold transition occurs when
$E_{W} > E_{F}$ or equivalently, using $a=c$, when $\Delta >
\Delta_{\text{F}}= (b/d) \lambda^2 /L$. Thus, this simple analysis has
recovered the correct scaling of $\Delta_{\text{F}}$ with respect to
$\lambda$ and $L$. As shown in Sec.~\ref{sec_transition}, the
prefactor $b/d$ turns out to be exactly 1.

\end{document}